\documentclass[manuscript]{acmart}

\usepackage{listings}
\usepackage{soul}
\usepackage[normalem]{ulem}
\usepackage{tabularx}
\usepackage{enumitem}

\AtBeginDocument{%
  \providecommand\BibTeX{{%
    \normalfont B\kern-0.5em{\scshape i\kern-0.25em b}\kern-0.8em\TeX}}}

\copyrightyear{2023} 
\acmYear{2023} 
\setcopyright{rightsretained} 
\acmConference[IUI '23]{28th International Conference on Intelligent User Interfaces}{March 27--31, 2023}{Sydney, NSW, Australia}
\acmBooktitle{28th International Conference on Intelligent User Interfaces (IUI '23), March 27--31, 2023, Sydney, NSW, Australia}
\acmDOI{10.1145/3581641.3584037}
\acmISBN{979-8-4007-0106-1/23/03}

\begin{document}

\title[The Programmer's Assistant]{The Programmer's Assistant: Conversational Interaction with a Large Language Model for Software Development}

%% This is a working title. We can certainly consider other %% suggestions!

\author{Steven I. Ross}
\orcid{0000-0002-2533-9946}
\affiliation{
    \institution{IBM Research AI}
    \city{Cambridge}
    \state{MA}
    \country{USA}
}
\email{steven_ross@us.ibm.com}

\author{Fernando Martinez}
\orcid{0000-0001-7172-4805}
\affiliation{
    \institution{IBM Argentina}
    \city{La Plata}
    \state{Buenos Aires}
    \country{Argentina}
}
\email{martferc@ar.ibm.com}

\author{Stephanie Houde}
\orcid{0000-0002-0246-2183}
\affiliation{
    \institution{IBM Research AI}
    \city{Cambridge}
    \state{MA}
    \country{USA}
}
\email{Stephanie.Houde@ibm.com}

\author{Michael Muller}
\orcid{0000-0001-7860-163X}
\affiliation{
    \institution{IBM Research AI}
    \city{Cambridge}
    \state{MA}
    \country{USA}
}
\email{michael_muller@us.ibm.com}

\author{Justin D. Weisz}
\orcid{0000-0003-2228-2398}
\affiliation{
    \institution{IBM Research AI}
    \city{Yorktown Heights}
    \state{NY}
    \country{USA}
}
\email{jweisz@us.ibm.com}

\renewcommand{\shortauthors}{Ross et al.}

\begin{abstract}
Large language models (LLMs) have recently been applied in software engineering to perform tasks such as translating code between programming languages, generating code from natural language, and autocompleting code as it is being written. When used within development tools, these systems typically treat each model invocation independently from all previous invocations, and only a specific limited functionality is exposed within the user interface. This approach to user interaction misses an opportunity for users to more deeply engage with the model by having the context of their previous interactions, as well as the context of their code, inform the model's responses. We developed a prototype system -- the Programmer's Assistant -- in order to explore the utility of conversational interactions grounded in code, as well as software engineers' receptiveness to the idea of \emph{conversing with}, rather than \emph{invoking}, a code-fluent LLM. Through an evaluation with 42 participants with varied levels of programming experience, we found that our system was capable of conducting extended, multi-turn discussions, and that it enabled additional knowledge and capabilities beyond code generation to emerge from the LLM. Despite skeptical initial expectations for conversational programming assistance, participants were impressed by the breadth of the assistant's capabilities, the quality of its responses, and its potential for improving their productivity. Our work demonstrates the unique potential of conversational interactions with LLMs for co-creative processes like software development.
\end{abstract}




%  \Description{Enjoying the baseball game from the %third-base

\maketitle

%%%%%%%%%%%%%%
% INTRODUCTION
%%%%%%%%%%%%%%

\section{Introduction}
Software development is a highly skilled task that requires knowledge, focus, and creativity~\cite{de-la-barra:creativity, Dekel:Interruptions}. Many techniques have been developed to enhance the productivity of software engineers, such as advanced code repositories~\cite{Spinellis:Git}, knowledge repositories~\cite{Jedlitschka:Repositories}, Q\&A sites~\cite{Abdalkareem:StackOverflow}, and pair programming practices~\cite{bryant:pair}. Collaborative software engineering is especially promising, given that professional software development is rarely a solo activity and relevant knowledge and expertise are typically distributed widely within an organization~\cite{Rejab:Distributed}. Many efforts have focused on incorporating collaborative technologies into software development environments (e.g.~\cite{cheng:Jazz, cook2005user, nordio2011collaborative, warner2017codepilot, anslow2013sourcevis}).

The pioneering work of Rich and Waters on \textit{The Programmer's Apprentice}~\cite{rich:apprentice} presented a novel concept of a knowledgeable automated assistant -- in effect, an artificial collaborative partner -- that could help software engineers with writing code, designing software systems, and creating requirements specifications. At the time, AI technologies and computing resources were not sufficient to fully implement their vision. In the intervening years, an increase in computational power, the availability of large corpora of language and code data, and the development of deep neural networks have made new approaches to achieving their goals worth exploring.

Recently, models leveraging the transformer architecture~\cite{Vaswani:Transformer}  have been developed to perform domain-specific software engineering tasks, such as translating code between languages~\cite{Roziere:Transcoder}, generating documentation for code~\cite{hu2020deep, iyer2016summarizing, wan2018improving, wang2022documentation}, and generating unit tests for code~\cite{tufano2020unit} (see \citet{talamadupula2021applied} and \citet{allamanis2018survey} for surveys). Recently developed foundation models -- large language models that can be adapted to multiple tasks and which exhibit emergent behaviors for which they have not been explicitly trained~\cite{bommasani2021opportunities} -- have also proven to be capable with source code. While the intent of training LLMs such as GPT-2~\cite{Radford:GPT2} and GPT-3~\cite{Brown:GPT3} was to give them mastery of natural language, it quickly became apparent that the presence of code in their training corpora had given them the ability to generate code based on natural language descriptions~\cite{metz:GPT-3}. The Codex model~\cite{Chen:Codex} was then produced by fine-tuning GPT-3 on a large corpus of source code data, leading to the development of Copilot~\cite{github:copilot}, a tool that helps software engineers by autocompleting code as it is being written. Experimentation with Copilot has shown its ability to perform additional tasks, such as explaining code, generating documentation, and translating code between languages~\cite{alvarado:Copilot-Labs}.

Although autocompletion interfaces are useful and valuable when the system can discern the developer's intent, there are many instances where that is insufficient. For example, the developer may have a good idea of what they want to do, but may be unclear on what functions, libraries, or even algorithms to employ. They may even have general programming questions that need to be answered before they are able to write any code.

In this paper, we seek to understand whether modern developments in code-fluent foundation models -- large language models that have been fine-tuned on source code data -- are sufficient to support a conversational agent that can act as an assistant in the software development process. We developed the Programmer's Assistant to explore the capabilities that conversational interaction could enable and the extent to which users would find conversational assistance with programming tasks desirable and useful.

We hypothesize that a conversational system may provide a flexible and natural means for interacting with a code-fluent LLM. Conversational interaction could enable users to pursue their questions in a multiple exchange dialog (as observed by \citet{Barke:Grounded}) that allows them to ask follow-up questions and refine their inquiries. A conversational programming assistant could ask the user clarifying or disambiguating questions to help it arrive at the best answer. It could also provide multiple types of assistance to the user beyond simply generating code snippets, such as engaging in general discussion of programming topics (e.g. \cite{catalan2021conversational, robe2022designing}) or helping users improve their programming skills (as observed in other studies of automating technologies~\cite{wang2019human}).

Our paper makes the following contributions to the IUI community:

\begin{itemize}
    \item We provide empirical evidence that a conversational programming assistant based on a state-of-the-art, code-fluent foundation model provides valuable assistance to software engineers in a myriad of ways: by answering general programming questions, by generating context-relevant code, by enabling the model to exhibit emergent behaviors, and by enabling users to ask follow-up questions that depend upon their conversational and code contexts.
    \item We show how different interaction models -- conversation, direct manipulation, and search -- provide complementary types of support to software engineers with tradeoffs between the user's focus and attention, the relevance of support to their code context, the provenance of that support, and their ability to ask follow-up questions.
    \item We motivate the need to further understand how to design human-centered AI systems that enhance the joint performance of the human-AI collaboration.
\end{itemize}

%%%%%%%%%%%%%%
% RELATED WORK
%%%%%%%%%%%%%%

\section{Related Work}
We discuss three areas of related work that have either motivated our study of conversational programming assistance or provided the technical foundations for it. We begin by briefly summarizing \citeauthor{rich:apprentice}' visionary work on the Programmer's Apprentice~\cite{rich:apprentice}, followed by summarizing work on code-fluent foundation models and human-centered evaluations of how these models impact software engineers' work. Finally, we discuss conversational interaction and how it might be employed to provide more flexible and sophisticated assistance to software engineers.

\subsection {The Programmer's Apprentice}
Our work is inspired by the vision laid out by \citet{rich:apprentice}, which describes an artificial agent that can act as an intelligent assistant for software engineers by providing advice, catching errors, and handling routine details throughout the software development process. The Programmer's Apprentice~\cite{rich:apprentice} relied on a knowledge base of ``clichés,'' which are formal, structured versions of what are known today as software design patterns~\cite{gamma:patterns}. It used a hybrid reasoning system capable of special-purpose reasoning based on frames and a plan calculus, along with general purpose logical reasoning. Although natural language interaction was envisioned, the original prototype implementation ultimately used a stylized command language. We view our work as a conceptual successor to the Programmer's Apprentice, as it enables the natural language interaction that the Programmer's Apprentice lacked.

\subsection{Code-fluent Foundation Models and Human-Centered Evaluations of Programming Assistance}

Generative models based on the transformer architecture~\cite{Vaswani:Transformer}  have recently been applied to the domain of software engineering. Code-fluent large language models are capable of generating code from natural language descriptions~\cite{xu2022ide}, translating code from one language to another~\cite{Roziere:Transcoder}, generating unit tests~\cite{tufano2020unit}, and even generating documentation for code~\cite{hu2020deep, iyer2016summarizing, wan2018improving, wang2022documentation}. These models are probabilistic systems, and as such, do not always produce perfect results (e.g. code that is free of syntax or logical errors). Nonetheless, \citet{Weisz:Perfection} found that software engineers are still interested in using such models in their work, and that the imperfect outputs of these models can even help them produce higher-quality code via human-AI collaboration~\cite{weisz2022better}.

New tools based on code-fluent LLMs are actively being developed. GitHub Copilot\footnote{\url{https://github.com/features/copilot}} is described as ``Your AI pair programmer.'' It is optimized for the code autocompletion use case: given a starting snippet such as a method's documentation, signature, or partial implementation, Copilot completes the implementation. Copilot is based on the OpenAI Codex model~\cite{Chen:Codex}, a 12 billion parameter version of GPT-3~\cite{Brown:GPT3, metz:GPT-3}, fine-tuned on code samples from 54 million public software repositories on GitHub. Empirical evaluations of this model have shown that, although the quality of its outputs is quite good, those outputs may still be problematic~\cite{nguyen2022empirical}. Echoing the results from \citet{weisz2022better}, human-centered evaluations of Copilot have found that it increases users' feelings of productivity~\cite{Ziegler:Productivity}, and that almost a third (27\%) of its proposed code completions were accepted by users. In a contrasting evaluation, \citet{Vaithilingam:Expectation} found that while most participants expressed a preference to use Copilot in their daily work, it did not necessarily improve their task completion times or success rates. Yet, in a study by \citet{kalliamvakou:copilot}, developers working with Copilot were able to implement a web server in Javascript 55\% faster than developers who did not use Copilot.

A grounded theory analysis of how programmers interact with Copilot~\cite{Barke:Grounded} found that their interactions varied depending upon whether they were accelerating tasks that they already knew how to do or if they were exploring solutions to problems that they were less sure about. Autocompletion was effective when developers were operating in ``acceleration mode'' and relied on the model to produce short completions that could be verified quickly. In ``exploration mode,'' however, the interaction was more awkward. Developers would communicate with Copilot by typing comments and seeing what Copilot generated in response. Then, they would modify their comments to explore other ways of prompting a response. Ultimately, the comments used to prompt the model would be deleted after the relevant code was generated, indicating that their value was largely in driving a back-and-forth, yet context free, dialog with the model to coerce it to produce the desired results through an iterative refinement process. In this paper, we fully commit to a context-aware conversational style of interaction with a code-fluent LLM and assess the value it provides to users.

\subsection{Conversational Interaction and Analysis}

\subsubsection{Conversational Interaction}
Using natural language to interact with technology has had a long research history~\cite{Adamopoulou:chatbots}, starting in the 1960s with pattern-matching approaches like Eliza~\cite{Weizenbaum:Eliza}, and continuing to today with state-of-the-art large language model-based conversational systems~\cite{Zaib:Survey} such as Meena~\cite{Adiwardana:Meena} and BlenderBot~\cite{Shuster:blenderbot}. These systems are intended to address the problem of open-domain dialog, with a goal of realistically engaging in conversation, but not particularly in a goal-directed or task-oriented manner.

Task-oriented chatbots are typically built with frameworks such as the Microsoft Bot Framework\footnote{https://dev.botframework.com/}, Google DialogFlow\footnote{https://cloud.google.com/dialogflow}, and IBM Watson Assistant\footnote{https://www.ibm.com/products/watson-assistant/artificial-intelligence}. They operate using pre-defined dialogue trees and use natural language processing to detect conversational intents and extract contextual entities. This structure enables the creation of special purpose, but fairly limited and rigid, conversational agents.

There have been several recent attempts to investigate conversational programming assistance. \citet{Kuttal:WOZ} conducted a Wizard of Oz study in which a pair programmer was replaced with a conversational agent, and they found that ``agents can act as effective pair programming partners.'' The PACT system~\cite{yadav:PACT} is a chatbot that assists programmers adjusting to new programming environments. PACT is structured as a discrete question-answering system based on a neural machine translation approach, but it doesn't maintain a conversational context.  

\subsubsection{Conversation Analysis}
Conversation is a form of interaction between people that enables robust communication. Conversation Analysis~\cite{Sacks:ConversationalAnalysis} is a method for understanding the natural structure of human conversational interaction. It catalogs different patterns of conversational acts and how they are utilized by interlocutors in order to attain a wide variety of goals. Recently, Conversation Analysis has been adapted to describe patterns of interactions between humans and artificial conversational agents in order to aid in the design of chatbots~\cite{Moore:ConversationalUX}. We apply techniques from Conversation Analysis in our study of conversational programming assistance.

%%%%%%%%%%%%%%%%%%%%%%%%
% PROGRAMMER'S ASSISTANT
%%%%%%%%%%%%%%%%%%%%%%%%

\section{The Programmer's Assistant}

\begin{figure*}[ht]
  \centering
  \includegraphics[width=\linewidth]{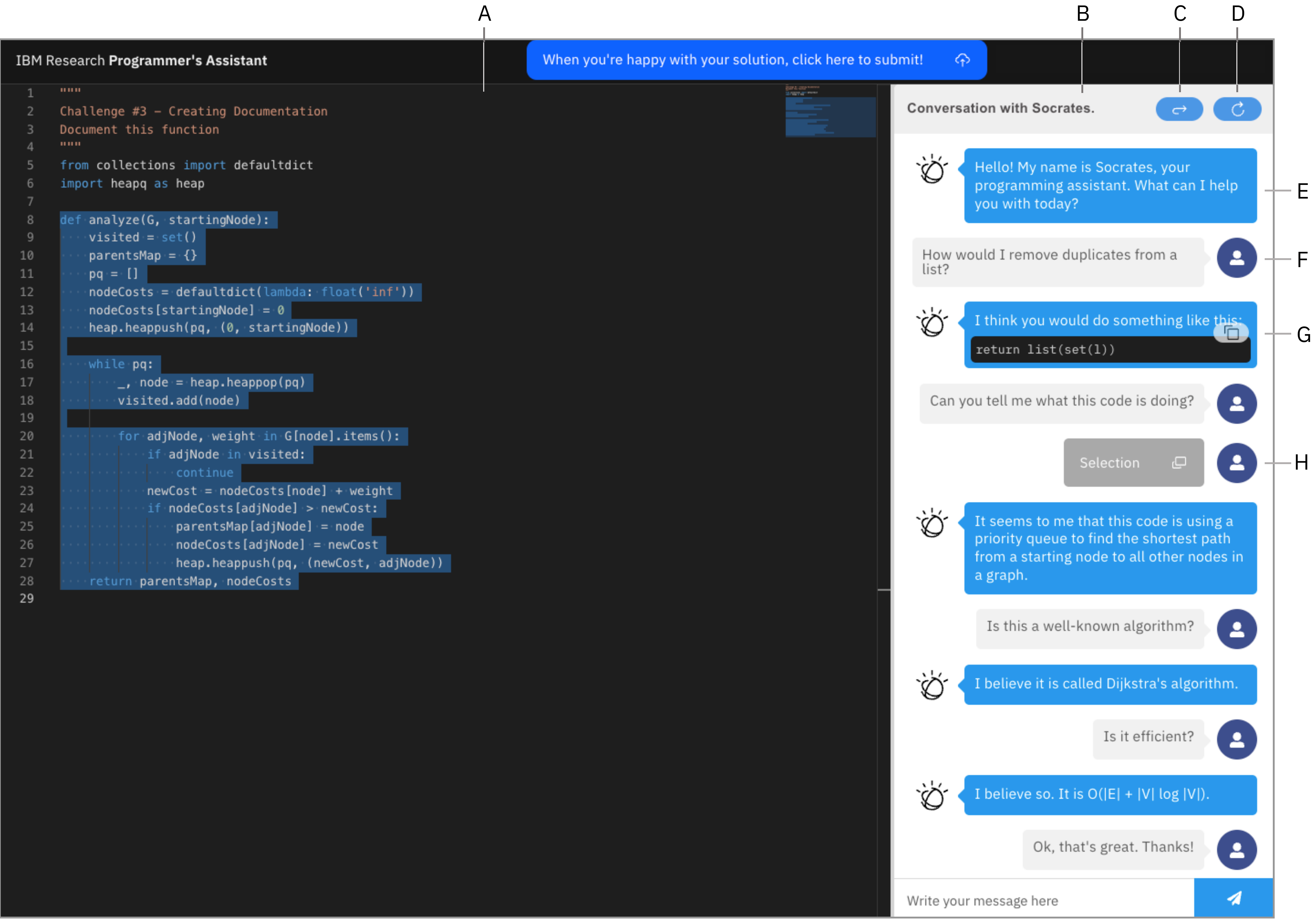}
  \caption{\textbf{The Programmer's Assistant.} The user interface provides a code editor on the left (A) and a chat pane on the right (B). The ``try again'' button (C) allows users to ask the assistant to generate an alternate response to the most recent question. The ``start over'' button (D) resets the conversational context for the assistant, but maintains the chat transcript in the UI. In this example, we show the assistant introduce itself to the user (E). Next, the user asks a general programming question (F), for which the assistant provides an inline code response (G). The user then asks a question about code selected in the editor (H), followed by a series of follow-up questions.}
  \label{fig:screen-shot}
\end{figure*}

In order to explore conversational programming assistance, we created a functional prototype system called \emph{The Programmer's Assistant}. Our prototype, shown in Figure~\ref{fig:screen-shot}, combines a code editor with a chat interface. The code editor was implemented using the Microsoft Monaco Editor\footnote{\url{https://microsoft.github.io/monaco-editor/}} embedded in a React wrapper\footnote{\url{https://www.npmjs.com/package/@monaco-editor/react}}. The chat user interface was implemented using the React-Chatbot-Kit\footnote{\url{https://fredrikoseberg.github.io/react-chatbot-kit-docs/}} framework. To drive the conversational interaction, we employed OpenAI's Codex model~\cite{Chen:Codex}, accessed through its web API.  

We developed our prototype as a lightweight coding environment in order to examine the user experience of interacting with a conversational assistant. Our work was exploratory in nature, and thus we did not have specific design goals for the prototype beyond integrating a code editor with a code-fluent LLM. We also did not attempt to target the prototype for a specific class of users (e.g. novices or experts) or use cases (e.g. writing code vs. learning a new programming language), as we wanted any value provided by conversational assistance to emerge from our user study. We also did not implement the ability to run or debug code in our prototype as we wanted to explore the nature of the conversational interaction rather than having users focus extensively on the production of working code.

When designing how users would interact with the Programmer's Assistant, we decided that it should be available on demand and not monitor the user's work in progress or give unsolicited suggestions or advice, in keeping with the conversational agent interaction model proposed by Ross et al.~\cite{Ross:Agent, Ross:Principles}. This approach was supported by feedback from prospective users who were concerned about the assistant providing criticism of unfinished efforts in progress or distracting them while they worked. Instead, we force initiative onto the user and only have the assistant respond to their requests. In this way, the assistant can provide help when requested without undesirable interruptions that can distract or interfere with the user's flow.

When a user interacts with the assistant, we keep track of their selection state in the code editor. If a user sends a message to the assistant without any code selected in the editor, then that message (along with the prior conversational context) is passed directly to the model. If a user sends a message to the assistant with new code selected in the editor (i.e. code that wasn't previously selected when they sent their last message), then that code is appended to the message before being communicated to the model.

The model may produce multiple types of responses to a user's message. We treat each type of response differently in the UI.

\begin{figure*}[ht]
  \centering
  \includegraphics[width=\linewidth]{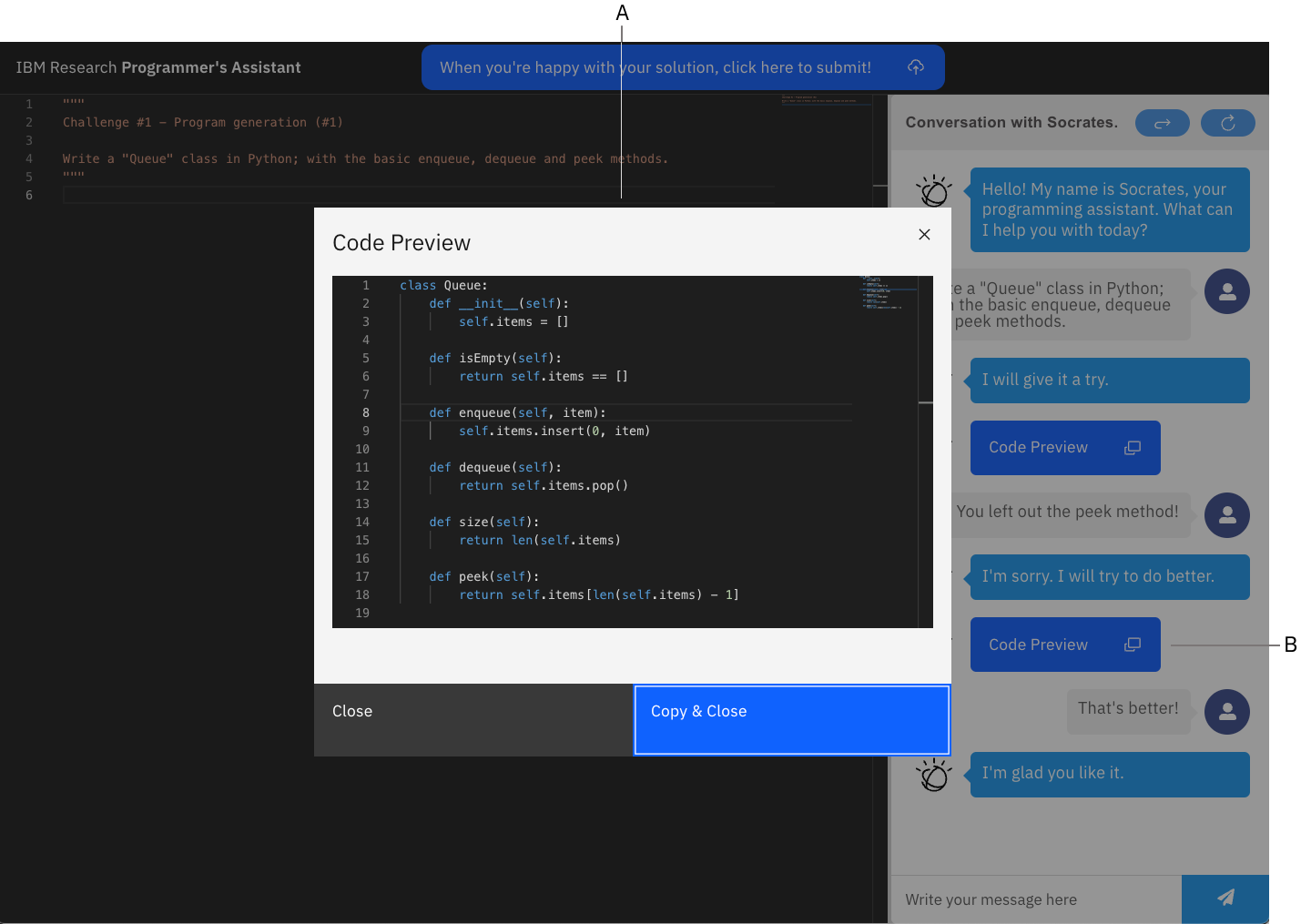}
  \caption{\textbf{Code Preview.} Long code responses (> 10 lines) are presented in a pop-up window (A), with a button to copy the code and close the window. This code can be re-accessed from the chat transcript by clicking the proxy icon (B).}
  \label{fig:screen-shot-long-snippet}
\end{figure*}

\begin{itemize}
    \item Responses that do not contain code are always rendered in the chat UI (Figure~\ref{fig:screen-shot}E).
    \item Responses containing short code snippets ($\leq$ 10 lines) are rendered inline in the chat UI (Figure~\ref{fig:screen-shot}G).
    \item Responses containing longer code snippets (> 10 lines) show the code in a pop-up window (Figure~\ref{fig:screen-shot-long-snippet}A), with a proxy entry in the chat transcript (Figure~\ref{fig:screen-shot-long-snippet}B) that allows users to re-display the code window after it has been closed. Non-code text in the response remains in the chat transcript.
\end{itemize}

The assistant never directly modifies the contents of the user's source code; rather, any code the user desires to transfer from the chat takes place via copy/paste.

Figure~\ref{fig:screen-shot} shows a screenshot of a real, sample conversation, in which the user asks a question that results in an inline response, then requests an explanation of some code in the editor, and then requests further elaboration. Figure~\ref{fig:screen-shot-long-snippet} shows an example conversation that resulted in the generation of a longer code sample, shown in a popup window. This example shows how the assistant produced an incomplete solution, followed by criticism from the user regarding the missing code, and resulting in an apology and the generation of a complete solution.

\subsection{Supporting Conversational Interaction}
We enabled Codex to conduct a conversational interaction by prompting it with a conversational transcript and a request to produce the next conversational turn. The prompt establishes a pattern of conversation between a user and a programming assistant named Socrates. It provides several examples of Socrates responding to general coding questions, generating code in response to a request, and accepting code as input. It establishes a convention for delimiting code in the conversation, making it easy to parse for display in the UI. It also establishes an interaction style for the assistant, directing it to be polite, eager, helpful, and humble, and to present its responses in a non-authoritative manner\footnote{The assistant's use of non-authoritative responses was encoded into the LLM prompt; output token probabilities from the LLM were not utilized to influence the assistant's response.}. Because of the possibility that the model might produce erroneous answers or incorrect code (as discussed in \citet{Weisz:Perfection}), we felt it was important that the assistant convey a sense of uncertainty to encourage users to not accept its results uncritically to avoid over-reliance (e.g. as observed in \citeauthor{moroz2022potential}'s study of Copilot~\cite{moroz2022potential}, and discussed more generally in \citet{ashktorab2021ai}) as well as automation bias~\cite{lyell2017automation, rajkomar2019machine, lou2021over}. We present the full text of the prompt used for the assistant in Appendix~\ref{appendix:prompt}.

\subsection{Architecture \& UI Design}

The Programmer's Assistant communicates with the Codex API via a proxy server that forwards requests from the React client. The proxy also rate-limits access to conform to the API's policy, and it logs UI events from the client (e.g. requests, responses, and UI interactions) in a back-end database. To address inconsistencies in the style or formatting of code generated by Codex, the proxy server reformats all code segments using the Black code formatter \footnote{\url{https://black.readthedocs.io/en/stable/}} before transmitting them to the client UI.

The client maintains the transcript of the ongoing conversation. Each time the user sends a message in the chat, the client constructs a new prompt for the model by concatenating the initial prompt, the chat transcript, and the user's new utterance, and makes a request for the model to complete the transcript. This completion request also specifies a stop sequence of tokens to prevent the model from generating both sides of the conversation (e.g. what the model thinks the user's next utterance might be after the assistant's response). Given the API's limitation on context length (4,096 tokens for both the prompt and model response), we silently ``forget'' older exchanges in the chat transcript when constructing the prompt to ensure that our completion request remains within bounds. Nonetheless, the entire conversational history remains visible to the user in the UI.

The client UI provides a loose coupling between the source code editor and the chat interface. Users can hide the chat pane when they wish to focus solely on their code, and re-engage with it when they desire assistance. Code selected in the editor is included in the conversation in order to couple the code context with the conversation. Easily-accessible buttons are provided in the UI to copy code responses from the assistant to the clipboard.

\subsection{Handling Model Limitations}
While developing the Programmer's Assistant, and in early pilot testing, we experienced some quirks and shortcomings of the model and our approach to using it for conversational interaction. One limitation stemmed from the fact that the model sometimes produced incorrect responses (e.g. code with syntax errors), incomplete responses (e.g. code that was missing functionality), irrelevant responses (e.g. responses not related to the user's question), or insubstantial responses (e.g. ``I don't know''). Because of the probabilistic nature of model inference, re-prompting the model would sometimes produce a more correct or appropriate response. Thus, we added the ability for users to ``try again,'' either by asking in the chat or by clicking a button in the UI (Figure~\ref{fig:screen-shot}{C}). This feature removes the assistant's last response from the context presented to the model and then re-invokes the model with an increased temperature\footnote{Temperature is a parameter in a generative model that specifies the amount of variation in the generation process. Higher temperatures result in greater variability in the model's output.}.

Although it is possible for transformer models such as Codex to produce multiple possible responses to a single prompt, we only request a single response in order to speed up response time as well as to preserve the token budget for conversational context. Thus, the ``try again'' feature provides an alternate way to produce a wider variety of responses.

During pilot testing, we noticed that the assistant sometimes happened to generate the same response to multiple, unrelated requests. In these cases, the assistant tended to get ``stuck'' in a pattern of repeating the same response and was unable to resume normal conversation. To avoid this problem, we \textit{automatically} execute a ``try again'' operation in the background when we see identical consecutive responses from the assistant.
 
Finally, we noticed that the accumulation of conversational context sometimes resulted in the assistant becoming fixated on some portion of the earlier conversation. For example, it might respond to a question with portions of the prompt or of earlier conversation, and become less responsive to newer requests. To address this issue, we introduced a ``start over'' feature, accessible via the chat or by clicking a button in the UI (Figure~\ref{fig:screen-shot}{D}), that resets the context to the original prompt, forgetting the rest of the conversational history.  We preserve the chat transcript in the UI, but delineate the break in the assistant's memory with an annotation in the chat transcript. These annotations are added both for ``try again'' and ``start over.''

\subsection{Sample Conversation}

We provide a real sample conversation with the Programmer's Assistant in Listing~\ref{lst:conversation}. This conversation begins with the assistant greeting the user (line 1). Next, the user asks a general Python programming question (line 4), to which the assistant responds with a non-authoritative remark (``I think...'') and a code snippet (line 9). The user next asks a follow-up question that depends on their previous question and the assistant's response (line 11), to which the assistant provides another code snippet (line 15), satisfying the user's request.

The user then switches topics and asks the assistant to write a Fibonacci function (line 17), and the assistant again responds with a non-authoritative remark (``I will give it a try,'' line 20) and a block of code. The user then asks how the function works (line 30) and the assistant provides an adequate description (line 32). Next, the user asks the assistant to re-implement the function in a different way (line 37), again leveraging the ability to ask follow-up questions. The assistant produces an alternative implementation that conforms to the user's request (line 41). The user follows up with a question that depends on multiple past utterances and responses in the chat transcript (line 47), and the assistant produces a relevant response (line 49). The conversation closes with the user thanking the assistant (line 53) and the assistant acknowledging their gratitude (line 55).

\begin{lstlisting}[numbers=left, basicstyle=\ttfamily\footnotesize, frame=single, caption={A conversation with the Programmer's Assistant. Code presented by the assistant is listed in \textbf{bold face}.},  xleftmargin=5mm, xrightmargin=1.5mm, label={lst:conversation}]
Socrates: Hello! My name is Socrates your programming 
          assistant.  What can I help you with today?

User:     How do I get a list of the keys in a 
          Python dictionary?

Socrates: I think you would do something like this:

          (*@ \textbf{list(my\_dict.keys())} @*)

User:     What if I just wanted to iterate over them?

Socrates: I think you would do something like this:
          
          (*@ \textbf{for key in my\_dict:} @*)

User:     Write a function that returns the nth value
          in the Fibonacci sequence.

Socrates: I will give it a try.

          (*@\textbf{def fib(n):} @*)
            (*@\textbf{if n == 0:} @*)
              (*@\textbf{return 0}@*)
            (*@\textbf{elif n == 1:}@*)
                (*@\textbf{return 1}@*)
            (*@\textbf{else:}@*)
                (*@\textbf{return fib(n - 1) + fib(n - 2)}@*)
            
User:     How does this work?

Socrates: The Fibonacci sequence is defined by the 
          recurrence relation F(n) = F(n-1) + F(n-2). 
          This function implements that recurrence
          relation.

User:     Can this be done without recursion?

Socrates: I think so. How about this?
          
          (*@\textbf{def fib(n):}@*)
            (*@\textbf{a,b = 0,1}@*)
            (*@\textbf{for i in range(n):}@*)
              (*@\textbf{a,b = b,a+b}@*)
            (*@\textbf{return} @*)
            
User:     Which version is better?

Socrates: I think the recursive version is more 
          elegant, but the iterative version is more
          efficient.

User:     Thanks!

Socrates: You're welcome!
\end{lstlisting}

%%%%%%%%%%%%%%%%%
% EMPIRICAL STUDY
%%%%%%%%%%%%%%%%%
\section{Empirical Study of Conversational Programming Assistance}
\label{sec:user-study}

We conducted an empirical user study of the Programmer's Assistant to assess whether conversational assistance provides value in a software engineering context\footnote{For historical context, we note that our study was completed before the public release of ChatGPT~\cite{openai2022chatgpt}, which has subsequently demonstrated the application of conversational assistance for programming tasks~\cite{glen2022chatgpt}.}. Our inquiry focused on the user experience and value of conversational interactions grounded in code. We therefore designed a qualitative study to investigate attitudes toward a conversational programming assistant: do people enjoy interacting conversationally, what kinds of questions do they ask, and how does the experience compare to other forms of programming support such as searching the web? We note that prior studies (e.g. \cite{weisz2022better, xu2022ide,  Ziegler:Productivity}) conducted quantitative examinations of the use of LLMs in code work; our study is akin to \citeauthor{Weisz:Perfection}'s qualitative examination of software engineers' attitudes toward working with models that may fail to produce working code~\cite{Weisz:Perfection}.

To address our questions, we deployed the Programmer's Assistant within our organization -- a global technology company -- and invited people to try it out and give us feedback on their experience. We invited people with varying levels of programming skill in order to obtain a wide range of feedback on the kinds of use cases for which the tool could provide assistance.

\subsection{Tasks}
\label{sec:user-study-tasks}

We set up the Programmer's Assistant as a playground environment that participants could try out with a few sample programming problems. We created a tutorial to orient participants to the assistant, its capabilities, and how to interact with it. We also created four programming challenges focused on writing code, documenting code, and writing tests for code. We designed these challenges to expose participants to a broad range of the assistant's capabilities. For each of these challenges, we explicitly did not evaluate metrics such as the participant's productivity, the quality of their solutions, or the time taken to produce them, as the focus of our study was to understand the utility of conversational interaction. We selected Python as the language used for the tutorial and challenges because of its general popularity~\cite{cass2022top} and the fact that it was well-supported by our underlying LLM~\cite{Chen:Codex}.

\subsubsection{Tutorial}
All participants were first introduced to the Programmer's Assistant through a tutorial. The tutorial walked each participant through 10 sample interactions to give them a feeling for what the assistant could do and how to interact with it. The tutorial demonstrated how to ask questions, how to request code to be generated, and how to evaluate existing code. It did not specifically cover how to generate documentation or unit tests. Tutorial instructions were provided within the code editor. We include the specific text used for the tutorial in Appendix~\ref{appendix:tutorial}.

\subsubsection{Programming Challenges}
After completing the tutorial, participants unlocked four programming challenges. Two of the challenges involved coding problems (writing a queue class and writing code to create a scatterplot of data in a CSV file), one involved documenting a given function (an implementation of a graph search algorithm), and one involved writing unit tests for a given function (computing the greatest common divisor of two arguments). Although the Programmer's Assistant was visible and available for use, we provided no specific requirement that it actually be used to complete the challenges.

After participants completed their solution to a challenge, they submitted it by clicking a button in the UI. The code editor used in the Programmer's Assistant was not a fully-functional IDE and did not provide syntax checking or the ability to run, test, or debug code. Due to these limitations, participants were asked to submit their solutions when they felt they had completed the challenge to their own satisfaction.

\subsection{Participants}
\label{sec:user-study-participants}
To recruit participants for our study, we posted internal advertisements in various communications channels focused on software engineering. Our advertisements stated that we were evaluating a conversational programming assistant, but were kept deliberately vague in order to minimize the impact on peoples' expectations of the experience.

Our advertisement yielded a pool of 140 potential participants. In order to recruit a diverse sample, we used a screening survey that asked about their job role, their familiarity with and recency of use of Python, and their availability to participate in our study. We accepted participants into the study on a rolling basis, selecting participants to capture a range of programming experiences and ensure balanced gender representation. We conducted periodic reviews to determine whether we were learning something new from each participant or if we had reached the point of saturation~\cite{ando2014achieving}. We stopped collecting data after running 42 participants as we were no longer observing any new behaviors or gleaning any new insights. The Programmer's Assistant implementation and configuration were held constant over the course of the study; no changes to the UI design or LLM prompt were made.

Our participants had the following self-identified characteristics:

\begin{itemize}
  \item \textbf{Job role:} 19 Software Engineers, 12 Researcher/Scientists, 3 Software Architects, 2 Data Scientists, 1 Machine Learning Engineer, 1 Systems Test Engineer, 1 Business Analyst, 1 Manager, 1 Marketer, and 1 Consultant.
  \item \textbf{Gender:} 21 Female, 19 Male, 1 Gender Variant / Non-conforming, and 1 Preferred not to say.
  \item \textbf{Python Experience:} 17 participants had 3+ years of Python experience, 11 had 1-3 years, 11 had less than 1 year, and 3 were not familiar with Python. 
   \item \textbf{Recency of Python Use:} 29 participants had written Python code within the past month, 4 within the past year, 5 within the past 5 years, and 4 had not written Python code within the past 5 years.
\end{itemize}

We provide full demographic information for individual participants in Appendix~\ref{appendix:participants}.

\subsection{Procedure}
\label{sec:user-study-procedure}
Participants completed the study on their own time, independently and without moderation. Each participant was provided with a web link to a pre-study survey that described the nature of the study and the tasks that they would be expected to perform. They were then directed to the Programmer's Assistant to complete the tutorial and the four programming challenges. When participants indicated they were finished with the challenges\footnote{We did not enforce that participants actually complete all of the challenges. Nevertheless, all participants but one did submit solutions to all of the challenges.}, they were directed to a final post-study survey. Complete sessions generally required about an hour of effort, though some participants spread their effort across a longer period of time and across multiple sessions. Participants were compensated for their time at a rate equivalent to US \$15/hr.

\subsection{Measures}
\label{sec:user-study-measures}
We collected a variety of data in our study from three sources:

\begin{enumerate}
    \item \textbf{Surveys}. We employed three surveys in the study: a pre-study survey to collect demographic information, a pre-task survey to gauge expectations of the conversational user experience, and a post-task survey to assess actual user experience. We describe these survey questions in the relevant context of our results, and we provide a complete listing of all survey instruments in Appendix~\ref{appendix:surveys}.
    \item \textbf{Event logs}. The Programmer's Assistant was instrumented to collect data on participants' usage. The event logs provided timestamped records of interaction events, including conversational exchanges, hiding/showing the assistant, use of the ``try again'' and ``start over'' features, and use of copy/paste.
    \item \textbf{Conversation logs}. From the event logs, we extracted conversational transcripts between each participant and the Programmer's Assistant.
\end{enumerate}

%%%%%%%%%
% RESULTS
%%%%%%%%%

\section{Results}

\subsection{Data \& Analysis}

We collected a wealth of data in our study: 126 survey responses from three surveys per participant, containing 296 written comments in open-ended survey questions, and 4,877 instances of 23 different types of UI events, including 1,699 conversational exchanges\footnote{We refer to a participant's utterance, followed by the assistant's response, as a conversational exchange.} in the event logs. We also compute, for each participant, counts or durations for 21 different metrics from the event logs.

In our analysis, we deliberately exclude the portion of our data collected during the tutorial exercise. We exclude this data because that activity was guided by the tutorial instructions, not by our participants' own initiative. Thus, our final sample consists of 3,172 events, including 968 conversational exchanges in the event logs; no survey data was excluded.

Our primary analysis of this data is qualitative, as our participants provided us with a rich source of interesting feedback and thought-provoking insights in their comments. Where applicable, we supplement this data with quantitative data from the survey and the event logs, as well as chat transcript data from the conversation logs. In this way, we triangulate~\cite{mackay1997hci} across our three data sources, using the open-ended survey data as a foundation. When we quote participants, either from their qualitative survey responses or the conversational transcripts, we reproduce their words exactly as typed, including typos, misspellings, grammatical errors, capitalization, and potential trigger words, and we only make minor clarifying edits where needed, delineated by square brackets.

In order to set the context for our analysis, we first describe how we used reflexive thematic analysis to analyze participants' responses to the open-ended survey questions. We then describe our analysis of the conversation logs and our development of a coding guide based on Conversation Analysis~\cite{Sacks:ConversationalAnalysis}, and specifically, \citeauthor{Moore:ConversationalUX}'s Natural Conversation Framework~\cite{Moore:ConversationalUX}.

\subsubsection{Thematic Analysis of Qualitative Survey Responses}

We conducted a reflexive thematic analysis to analyze the responses to our seven open-ended survey questions. We followed the process described by \citet{braun2022reflexive} in which researchers immerse themselves in the data, generate codes for material that seems interesting, and then iteratively group and refine codes through collaborative discussion in order to identify higher-level themes. Initially, four authors performed open-coding on the open-ended survey responses. Through discussion, these codes were grouped and consolidated into a single set, which were then re-applied to the data by two authors. After another round of discussion, these authors identified a set of 12 higher-level themes. Some themes had clear parallels to quantitative survey questions or event log data, and thus represented clear instances where we were able to triangulate across data sources. Other themes surprised us. We structure our presentation of the results based on these 12 themes, grouped into three different aspects of the user experience: expectations and experience, utility of conversational assistance, and patterns of interaction and mental models.

\subsubsection{Conversation Analysis via the Natural Conversation Framework}

\begin{table*}[ht]
    \centering
    \begin{tabularx}{\linewidth}{llX}
        \toprule
        \textbf{Interlocutor} & \textbf{Orientation} & \textbf{Codes} \\
        \midrule
        Human & Social & \textbf{Acknowledgment}, \textbf{Apology}, Criticism, \textbf{Expression of Gratitude}, \textbf{Farewell}, \textbf{Greeting}, Politeness, Praise, \textbf{Self Identification}, Small Talk, \textbf{Welfare Check}, \textbf{Welfare Report} \\
              & Task & Asks Question, Asserts Information, Capability Check, Command of Action, Expression of Desire, Identifies Error, Request for Action, Requests Elaboration, Requests Explanation \\
              & Meta / UI & Chat Context Required, \textbf{Confirmation}, Copy, Copy (Extraneous), Erroneous Input, Includes Selection, Includes Extraneous Selection, Missing Selection, Paste, Paste (Extraneous), Pasted Code in Chat, Spelling Error, Start Over, Try Again \\
        \midrule
        Assistant & & Appears Fixated, Claims Ignorance, Didn’t Understand, Grants Request (Complete), Grants Request (Incomplete), Offers Help, Provided Wrong Answer, Requests Details, Requests Paraphrase, Response Includes Code, Spews Garbage \\
        \bottomrule
    \end{tabularx}
    \caption{\textbf{Event log codebook.} Our codebook contained 46 unique codes, applied separately to participant utterances (Human) and assistant responses (Assistant). Codes in \textbf{bold} were applied to both participant and assistant responses. Human codes were classified as demonstrating either a social or task orientation to the assistant.}
    \label{tab:event-codebook}
\end{table*}

In order to understand the content and structure of the conversations that took place between our participants and the Programmer's Assistant, we turned to the Natural Conversation Framework~\cite{Moore:ConversationalUX} (NCF). 
We developed a codebook for the event logs, beginning with 21 different categories of utterances from the NCF. Nine NCF categories -- Acknowledgment, Apology, Confirmation, Expression of Gratitude, Farewell, Greeting, Self-Identification, Welfare Check, and Welfare Report -- appeared twice in our codebook to distinguish cases in which the utterance was made by the human participant vs. the assistant. Other NCF categories were split to provide nuanced detail about the interaction; for example, we distinguished three different kinds of NCF requests, depending upon whether they were stated as Requests for Action (e.g. ``Would you...''), Commands of Action (e.g. ``Write a function that...''), or Expressions of Desire (e.g. ``I want...''). We also added 18 additional codes to identify meta-information such as utterances that included code, utterances that referenced selected code, utterances that implicitly or explicitly referenced earlier portions of the conversation, or non-verbal UI activities such as copies, pastes, and invocations of ``try again'' and ``start over.'' Finally, we classified a subset of the human-applied codes based on whether they represented a participant's task or social orientation toward the assistant. We list our codes in Table~\ref{tab:event-codebook}, but note that not all of them ended up being relevant to our analysis.

When coding conversational data, we applied individual codes at the level of each conversational utterance. We allowed multiple codes to be applied to each utterance to account for utterances that performed multiple functions (e.g. greeting and self-identification). In order to ensure consistency in how our codebook was applied, two authors coded a 10\% sample of the 968 conversational exchanges, achieving a satisfactory level of inter-rater reliability (Krippendorf's $\alpha = 0.77$, where agreement was conservatively defined as having all of the same codes applied to both utterances in a conversational exchange).

\subsection{Expectations and Experience}
Pilot testing of the Programmer's Assistant suggested that software engineers would be skeptical of a conversational programming assistant and its ability to provide useful assistance. Our study revealed that, for most participants, their actual experience after using the tool was better than they had anticipated. Participants were surprised at the quality of the assistant's responses and they appreciated how its integration with the code editor reduced the amount of context switching they needed to do in the UI. Some participants struggled with the code selection feature, although others appreciated the ability to ask questions related to selected code.

\subsubsection{Usage}
\label{sec:usage}

All of our participants engaged with the Programmer's Assistant while working on the challenges, despite there being no requirement to do so. Forty-one participants submitted solutions to all four challenges, and one participant, P14, only submitted solutions for one of the four challenges. Participants spent an average of 68 minutes engaged with the assistant, as measured by the amount of time the Programmer's Assistant window was in focus.

Participants made an average of 23.0 utterances (SD = 15.1 utterances) to the assistant. On average, 6.2 of their utterances (SD = 4.3 utterances) contained a code selection. The average latency per request\footnote{This time includes additional time added by our proxy server to ensure our conformance to the API rate limitation.} was 6.7 seconds (SD = 3.1 seconds). 

We saw a 66.3\% rate of acceptance of generated code, where we considered code to be accepted if the participant performed a copy immediately after the code was generated.  This acceptance rate is much higher than the 27\% acceptance rate reported for Copilot~\cite{Ziegler:Productivity}. We believe one reason we observed a higher acceptance rate is because Copilot's completion suggestions are generated proactively, whereas the Programmer's Assistant's suggestions are generated upon request. When copying generated code from the assistant, participants most often copied the entirety of the generated code, and only in 5.8\% of cases did they copy a smaller portion of it.

\subsubsection{User Experience Expectations \& Changed Attitudes}
Prior to running our study, we had reason to believe that participants would be skeptical of a conversational programming assistant. Before developing the Programmer's Assistant, we showed potential users mockups of a program editor with an integrated chatbot feature. These prototypes elicited uniformly negative reactions. People told us about their frustrating experiences with conventional chatbots and raised doubts about the knowledge, capabilities, and value of a conversational programming assistant. This skepticism motivated us to develop the Programmer's Assistant in order to evaluate whether the conversational experience, as powered by a state-of-the-art code-fluent LLM, would be better than people had anticipated. During pilot testing, we received feedback that the Programmer's Assistant provided a much better conversational experience compared to testers' previous experiences with chatbots. Thus, in designing our study, we felt it important to first gauge participants' expectations of a conversational interaction around code, and then measure their experience after the fact.

We developed a short inventory of six scale items to measure user experience of code work\footnote{Our scale items were modeled from scales published in \citet[Table 9 -- AI Support]{weisz2022better} that measured constructs including ease of use (item 3), response quality (item 1), the production of higher-quality code (item 5), and the ability to write code more rapidly (item 4). We added additional items to cover the constructs of request understanding and enjoyment, and we cast all items on a 4-point scale of extent.}. The scale was administered twice: once before participants were exposed to the Programmer's Assistant (but after they had been briefed that they would interact with an AI chatbot), and once after completing the programming challenges. The items were presented with the appropriate tense: Do you expect (Did you find that) the Programmer's Assistant: (a) will be (was) easy to use; (b) will understand (understood) your requests; (c) will provide (provided) high quality responses; (d) will help (helped) you to write better code; (e) will help (helped) you to write code more quickly; (f) will be (was) enjoyable to use. Each item was rated on a 4-point scale of extent: Not at all (1), A little (2), Somewhat (3), A great deal (4).

A factor analysis revealed the items on this scale measured a single construct, which we identify as user experience (Cronbach's $\alpha$ = 0.87). Thus, we computed two scores of user experience (UX) for each participant: a pre-task UX score computed as the average of their six pre-task expectation scale responses, and a post-task UX score computed as the average of their six post-task experience scale responses.

We found that participants had lower initial expectations for their experience with a conversational programming assistant (pre-task UX M (SD) = 3.0 (0.62) of 4) than their experience actually was (post-task UX M (SD) = 3.6 (0.32) of 4). A paired sample t-test shows that this difference was significant, $t(41) = 5.94$, $p < .001$, Cohen's $d = 0.92$ (large). Measured another way, 32 participants (76.2\%) had post-task UX ratings that were higher than their pre-task expectations, demonstrating a significant shift in attitudes toward conversational programming assistance.

However, the UX ratings alone fail to capture participants' nuanced expectations of the assistant and the reasons for their shifted attitudes after using it. Participants expressed a variety of expectations of the assistant before using it, including that it would be easy to use (P30) and produce correct responses (P30), understand the problem and what is being asked of it (P8, P9, P11), not interfere with their flow state (P5), produce imperfect or questionable outputs (P6, P21), improve with feedback (P31), provide generic and unhelpful answers (P17) or only answer basic questions (P40), and produce responses quickly (P40).

P17 expected \emph{``to be frustrated very quickly and that what I'd think would be relatively common questions would be responded to with generic, unhelpful answers.''} P6 explained,  \emph{``I didn't have very good experiences with chatbots. I think I'll need to spend more time in reviewing and fixing the suggestions than in writing the code myself from scratch.''} P11 had a more balanced view, that \emph{``It'll do some tasks really well, but others will not be as reliable.''}

After interacting with the Programmer's Assistant, many participants commented on how the experience was better than they anticipated, because it \emph{``seemed to be able to handle complex issues''} (P10) and \emph{``was a great help''} (P8). P20 felt it was \emph{``incredible!''} P6 and P17, who were both initially skeptical, reported having a positive experience. For P6, \emph{``It absolutely exceeded all my expectations, in all aspects that I could have imagined and more!''} P17 provided a more quantitative assessment: \emph{``Initial expectations:  3 Actual:  9.5.''} P38 was emphatic in their evaluation: \emph{``I was blown away how well it allowing me to structure how I want the code to look and work and just giving me the thing I asked for.''}

Many participants described a sense of surprise in their experiences. P9 was surprised by how well it understood their requests:

\begin{quote}
    \emph{``I was surprised at how well the Programmer Assistant was able to understand my requests and generate good code/documentation/tests. It understood major concepts and was able to explain it to me in a clear way, and it was also able to understand and write functional code. It even was able to help me review my answer. I was also surprised at how well it could understand the context of what I was asking in follow-up questions when I did not specify exactly what I was talking about, but rather referencing our prior conversation (such as, `what does that mean').'' (P9)}
\end{quote}

Similarly, P6 was surprised that they liked the conversational interaction when they expected that they wouldn't:

\begin{quote}
    \emph{``I though[t] I wouldn't like the chatbot interaction and that I would prefer something like the tool I've seen in those demos [of Copilot]. But surprisingly, after using the chatbot (and seeing the results: easy to use, it understands well, I felt it like a partner) I like this kind of help.'' (P6)}
\end{quote}

\subsubsection{Quality of Assistant’s Responses}
\label{sec:quality-of-assistants-responses}

In order to gauge the quality of responses produced by the Programmer's Assistant, we examined the 910 task-oriented requests made by participants in the study. For the vast majority (80.2\%), the assistant produced a correct response (Grants Request (Complete)); in other cases, the assistant's response was incorrect (9.6\%; Provided Wrong Answer), correct but incomplete (4.4\%; Grants Request (Incomplete)), or the assistant didn't understand (3.4\%; Didn't Understand), claimed ignorance of the subject (1.5\%; Claims Ignorance), or produced another type of response (0.9\%; Appears Fixated, Spews Garbage).

Participants also reported experiencing this variability in the quality of the assistant's responses. Some participants described how the assistant provided \emph{``detailed answers''} (P17) and \emph{``high quality outputs''} (P18) that were \emph{``surprisingly good''} (P2). P6 felt it was \emph{``incredible to see the quality of the responses,''} and P3 even explored the assistant's capabilities outside the scope of the challenges and found that it could handle those as well:

\begin{quote}
    \emph{``It was surprising the quality of the code and the ability to answer all my questions correctly. Although I think the challenges may be biased towards what the Assistant is able to do, it was a great experience because I asked many other things and it was able to answer correctly.'' (P3)}
\end{quote}

Of course, the Programmer's Assistant wasn't perfect, and some participants did run into issues. For P35, \emph{``The documentation generation did not perform very well.''} P16 questioned the accuracy of the knowledge encoded in the model: \emph{``Does the model need to be updated? It said latest python version is 3.7 but google says it's 3.10.''} In some instances, participants needed to ask their question multiple times to get a good response: \emph{``you need to ask many times if you want to get an answer and also a detailed explanation''} (P3). P27 felt, \emph{``it was annoying when I asked it to try again and it would give me the same response.''} P22 struggled because, \emph{``It didn't seem to handle multiple sentences well.''}

P28 perhaps offered the most scathing criticism, that, \emph{``It makes mistakes often enough to be not very practical.''} However, despite the production of poorer-quality responses, other participants felt that the assistant was still helpful. P36 reported that, \emph{``Only minor tweaks were normally needed to correct any issues.''} Similarly, P38 described how the assistant wasn't able to completely solve their problem, but provided a useful start:

\begin{quote}
    \emph{``There was only one hickup I noticed where when I asked it to memoize fibonacci it couldn't, but it dropped the building blocks on my lap for me to finish so that was fine, that was like minutes of effort on my part.'' (P38)}  
\end{quote}

\subsubsection{UI Design \& Affordances}
Participants made many comments on our specific UI design and the affordances provided (or not provided) in our chat-augmented editor. Overall, the integration between the chat pane and the code editor was \emph{``very good''} (P23), with a \emph{``nice interface between the code pane and the assistant pane''} (P17) that \emph{``makes it really convenient''} (P35).

Prior research by \citet{brandt2010example} has shown how keeping developers focused in their IDE improves productivity, and our participants expressed similar sentiments. P40 remarked, \emph{``It allows me to stay in one browser window/tab!''} and P12 hinted at how the interface might preserve their flow state by \emph{``prevent[ing] me from getting distracted when looking into an issue in another tab.''} 

Some aspects of our user interface were confusing to participants, such as the mechanism for selecting code to be included in the conversational context. P7 remarked, \emph{``It's was a little confusing doing the selection part for it to tell me what a function does, but... it gave me code that was insanely easy to copy and paste.''} Other participants appreciated the code selection mechanism, such as P11: \emph{``I enjoyed the code selection feature, and found that very easy to use.''} In the event logs, we identified 20 instances in which a participant unintentionally included selected code in the conversation when it wasn't needed (Includes Extraneous Selection), 12 instances in which a code selection was omitted when it was needed to provide context for the question (Missing Selection), and 16 instances in which a participant copy/pasted code directly into the chat rather than selecting it in the editor (Pasted Code in Chat). Although these cases represent a small fraction of the 227 instances in which a code selection was required and included in the conversation (Includes Selection), their presence does indicate that more attention is needed to the interaction design of code selection.

Another issue regarded the awareness of the ``try again'' and ``start over'' features. The ``try again'' feature was only used by 14 participants, who used it a total of 63 times over the course of the study. Some participants used it specifically when they got an answer which they saw as clearly wrong, while others used it to get a variety of possible answers before proceeding. The ``start over'' feature was used even less, by 5 participants who used it a total of 6 times. Despite our effort to surface these conversational features in the UI via shortcut buttons, they may not have been sufficiently noticeable or salient: \emph{``The `try again' button is not so reachable, often times I forgot it exists''} (P23). By contrast, at least one participant was successful with these features:

\begin{quote}
    \emph{``at some point it had issue with challenge 3 and I had to start over. Just asking `try again' was not enough and I was getting always the same (wrong and not related) answer. starting again solved the issue!'' (P20)}
\end{quote}

\subsection{Utility of Conversational Assistance}
Our next set of themes concerns the utility provided by conversational programming assistance. Participants felt the assistant was highly valuable and desired to use it in their own work. They felt it would be most helpful for smaller or narrowly-scoped tasks, but able to provide a wide variety of types of assistance. The fact that the interaction model was conversational and grounded in code were valuable aspects, as was the ability for the assistant to bolster users' learning about programming topics through that interaction. Participants did question whether they could trust and rely upon the assistant's responses, echoing a similar theme discussed in \citet{Weisz:Perfection}.

\subsubsection{Value \& Appropriate Tasks}
Participants rated the value of the Programmer's Assistant highly (M (SD) = 8.6 (1.4) of 10). Many participants asked questions such as, \emph{``Can I have it in my editor please?''} (P15), or made comments that, \emph{``I would enjoy using it in the future''} (P36), \emph{``I would love to be able to... have access to it for my coding''} (P37), and \emph{``I'd love to use this tool as part of my usual programming workflow if I could!''} (P39). Some of the reasons why participants found it valuable are because it \emph{``help[s] me remember how to do things in certain languages that normally I would just Google''} (P9) and \emph{``It helps me to avoid silly syntax errors and can when I cannot remember exact function/method names and required arguments''} (P40). We did not observe any differences in value ratings based on participants' familiarity with or recency of using Python.

Participants described a wide variety of tasks for which they felt the assistant would be useful. These tasks included \emph{``ordinary''} (P23), \emph{``simpler''} (P2), and \emph{``small, repetitive''} (P4) tasks such as \emph{``quick lookups''}  (P25) for \emph{``short chunks of code''} (P11) or for \emph{``narrowed questions''} (P26). Participants also felt the assistant was useful for \emph{``small containable novel algorithms''} (P38) and \emph{``little coding problems''} (P4).

Several kinds of task assistance were reported as being valuable, such as explaining code (P31), implementing business logic in a UI (P38), understanding what code does (P19, P37), and recalling language syntax, method names, and arguments (P12, P15, P20, P40, P42). P27 felt that the assistant was \emph{``More helpful when recognizing a specific well known algorithm but not things you make yourself.''}

Participants also made recommendations for how to increase the value of the Programmer's Assistant. P38 suggested, \emph{``What would blow me away though is if it's able to help with what I do most often which is to integrate, refactor and iterate on an existing system.''} P16, P26, and P38 all desired more information on the data sources used to produce the assistant's responses. P9 requested to \emph{``Have the Programmer's Assistant examine your code and make proactive suggestions for improving it in the chat.''} P36 requested the same, but cautioned that, \emph{``Care would need to be taken to avoid becoming an annoyance or disrupting the flow of a coding session.''}

In the post-task survey, we probed participants on how certain changes to the Programmer's Assistant would either decrease, increase, or result in no change to its value. Over 75\% of participants felt that the assistant would be more valuable if it operated in a proactive manner, either by making improvement suggestions in the chat or as comments directly in the code. Similarly, 78.6\% of participants felt that having more buttons in the UI for common features such as explaining or documenting code would make the tool more valuable.

\subsubsection{Conversational Interactions Grounded in Code} 
\label{sec:conversational-interactions-grounded-in-code}
One of the challenges in interpreting participants' comments about the utility of the Programmer's Assistant was in disentangling the extent to which value was derived from the quality of the underlying model versus the integration of conversation in a code context. Indeed, participants felt that the chat interaction was valuable: 69.0\% of participants felt that eliminating the conversational interaction and making the assistant behave more like web search would decrease its value. Further, our analysis of the conversation transcripts revealed that 42\% of the 910 task-oriented utterances from participants required historical conversational context (Chat Context Required) in order to be correctly interpreted. Thus, we observe that behaviorally, participants did rely on conversational context in their interactions.

In the post-task survey, 83\% of participants rated the importance of the ability to ask follow-up questions as being ``somewhat'' or ``a great deal.'' Several participants specifically commented on the value of this conversational context. P39 remarked, \emph{``I absolutely loved how you can straight up ask follow-up questions to the Programmers' Assistant without having to reiterate the original topic/question.''} P15 expressed a similar sentiment, saying, \emph{``I think the conversational context was someone helpful, just in communicating that it's a running conversation where my context is remembered.''} P9 provided a similar analysis:

\begin{quote}
	\emph{``This tool was so helpful at answering questions I had about the code in the context of the code I am working on... I was also impressed with how well it was able to remember the context of our conversation, especially when I asked vague follow-up questions.''} (P9)
\end{quote}

In addition, some participants identified how a conversational interaction grounded in code was useful, \emph{``because I think to `understand' the dev context could be VERY important''} (P31). In fact, 24.9\% of task-oriented utterances included a relevant code selection (Includes Selection), showing that participants valued this ability.   

Contrasting with these participants, P18 felt that interacting with the assistant conversationally was tedious, and they employed a more direct approach:

\begin{quote}
	\emph{``I really like the PA. But, I didn't converse with it like a chat bot. I often told it what to do (`Document this code.') as opposed to asking it what to do (`How do I document this code?'). Talking to it the way that was suggested in the tutorial seemed overly verbose/tedious.''} (P18)
\end{quote}

Despite these individual differences in interaction preferences, P39 envisioned that both interaction styles could be supported in the tool:

\begin{quote}
	\emph{``I think both options should exist: people should be able to input their queries like a search bar AND also give their question as if in conversation.''} (P39)
\end{quote}

\subsubsection{Learning Effects}
\label{sec:learning-effects}

One specific benefit of the Programmer's Assistant identified by participants is its ability to help people improve their programming skills and reinforce knowledge gaps. For example, it can help users \emph{``remember how to do things in certain languages... such as, when I am using a language I haven't used in a while''} (P9). The assistant can also serve as an memory aid, such as when \emph{``I use a lot of libraries that I don't always remember all of the functions''} (P15). Similarly, P31 said, \emph{``No matter how good you're as a developer, you can't (humanly) remember all the API of hundreds of libs or new languages... I'd learn new dev lang and new lib/frameworks faster.''}

P39 felt the assistant \emph{``is perfect for programmers of all levels,''} and P1 felt it could help them rapidly improve their Python skills:

\begin{quote}
	\emph{``I have wanted to learn python... The main concern how much time spent learning is needed before I could actually get some value out of learning python. I have a feeling this would cut that time down from weeks to a day or so.''} (P1)
\end{quote}

P39 also identified the fact that, because the interactions with the assistant are conversational, it forces people to learn how to communicate to others about their code:

\begin{quote}
	\emph{``The conversation aspect promotes proper communication, which would really stand to benefit budding programmers if they want to learn how to explain concepts more fluently in the future to their colleagues.''} (P39)
\end{quote}

Conversely, P36 suggested that over-reliance on programming assistance might have a detrimental effect to one's learning:

\begin{quote}
	\emph{``It's definitely a huge time saver, but over-reliance on it may cause new developers to skip learning the reference material themselves and discovering new things and sparking new ideas.''} (P36)
\end{quote}

\subsubsection{Trust}
Many participants raised questions about whether they could trust the responses provided by the Programmer's Assistant. P21 asked this question most directly: \emph{``will the code be correct, safe, efficient?''} Other participants raised similar questions, such as, \emph{``I'm wondering how it validates it's answers, if it can be trusted to always give a working answer''} (P10), and \emph{``Sometimes lack of source and context may raise doubts in the mind of the programmer''} (P16).

These issues of trust were exacerbated by the fact that the Programmer's Assistant did not allow participants to actually run their code. Because of this limitation, participants had to rely on their own knowledge to judge the correctness of the assistant's responses. P19 asserted, \emph{``There is no way to evaluate if the Programmer's assistant is giving you the right advise or not other than your own knowledge,''} and P9 concurred: \emph{``I had to trust that it was correct (and use my own prior knowledge).''}

P18 described the potential consequences of allowing the assistant to write code for them:

\begin{quote}
	\emph{``The only thing that made me nervous was that it could have introduced a bug that wasn't immediately apparent. And given I didn't write the code, I could have easily glossed over a mistake when reviewing it. Especially if it is also the one writing the test cases.''} (P18)
\end{quote}

Despite our efforts to make the Programmer’s Assistant respond in non-authoritative ways, we did observe participants sometimes uncritically accept generated results that were clearly wrong or incomplete. Thus, we did find behavioral evidence for over-reliance.

\begin{lstlisting}[numbers=left, basicstyle=\ttfamily\footnotesize, frame=single, caption={Building trust through explanations and justifications}, xleftmargin=5mm, xrightmargin=1mm, label={lst:justification}]
P9:       what is a unit test?

Socrates: A unit test is a test of a single function 
          or method.  It is typically written by the
          programmer who wrote the code.

P9:		
          (*@\textbf{def euclid(a, b):}@*)
             (*@\textbf{while b != 0:}@*)
                (*@\textbf{(a, b) = (b, a \% b)}@*)
             (*@\textbf{return a}@*)

          how would I write a unit test for this code?

[...]

Socrates: I think you would do something like this:

          (*@\textbf{def test\_euclid():}@*)
            (*@\textbf{assert euclid(2, 4) == 2}@*)
            (*@\textbf{assert euclid(6, 9) == 3}@*)
            (*@\textbf{assert euclid(17, 12) == 1}@*)
            (*@\textbf{assert euclid(4, 4) == 4}@*)

P9:       what is a euclid function

Socrates: The euclid function is a function that
          computes the greatest common divisor of two
          numbers.

P9:       is this a good unit test for a euclid 
          function

Socrates: I believe that it is a good unit test for
          a euclid function.

P9: 	  why

Socrates: It seems to me that it tests the edge
          cases of a and b being equal, a being 
          greater than b, and b being greater than a.

[...]

P9: 	  thanks Socrates this was helpful

Socrates: Always happy to help.
\end{lstlisting}

One way to address trust issues is for the assistant to provide further explanations and justifications that can calibrate a user's confidence in the assistant's responses. Such explanations could be requested conversationally, though most participants did not attempt to do so. One participant (P9) did ask for such explanations, and we show a summary of their transcript in Listing~\ref{lst:justification}. In this instance, P9 asked for a definition of a unit test (line 1), an explanation of the code being tested (line 25), and justifications of the quality of the unit test (lines 31\& 37). Thus, we observe that the assistant is capable of producing explanations and justifications when asked.

\subsection{Patterns of Interaction and Mental Models}

Participants interacted with the assistant in a variety of ways with two main patterns of usage standing out: (1) invoking the assistant to solve the entire programming challenge, and (2) breaking the challenge down into a set of smaller tasks and invoking the assistant's help for each. There were no clear differences in how participants with differing Python experience approached the tasks.

Participants' mental models of the assistant also varied. Although participants strongly saw the role of the assistant as being a tool, their behaviors revealed that in many cases, they actually treated it as a social agent. In addition, participants ascribed various mental capacities to the assistant, such as having the ability to understand, compute, and learn.

Participants felt the assistant changed the nature of their work process. For some participants, it enabled them to focus on the higher-level aspects of development because the assistant handled lower-level details or provided partial solutions for them to build upon. Many participants felt the assistant sped up their work and helped them remain focused on their tasks.

Finally, participants drew comparisons between the Programmer's Assistant with other forms of programming support such as Copilot and web search. They felt that the conversational style of interaction enabled them to discover new, emergent behaviors from the model that were unavailable from Copilot's focus on code autocompletion. They also felt that the examples provided by the assistant were more readily usable within their own code compared to browsing for answers within search results, speeding up the coding process. However, some participants advocated for a balanced approach to the design of programming assistance tools by incorporating multiple modes of interaction rather than fixating on a single one.

\subsubsection{Interaction Styles and Assistant Role}
We observed that participants interacted with the Programmer's Assistant in strikingly different ways. Some participants would present the entire challenge description to the assistant and then work with the results it produced. Other participants approached the programming challenges in a piecemeal fashion, breaking them apart into a set of smaller tasks, then invoking the assistant to aid with each one.

Experience with Python was not a determinant of how participants approached the programming challenges, but it did seem to impact how participants interacted with the assistant. Less experienced participants tended to ask the assistant basic questions such as, \emph{``What is a unit test''} (P29, not familiar with Python) and \emph{``how do I document a function?''} (P27, < 1 year of experience). More experienced participants made detailed requests about specific Python libraries or algorithms, such as, \emph{``given a pandas dataframe with two columns `Date' and `Sales' please use matplotlib to draw me a scatterplot''} (P38, 3+ years of experience) and \emph{``implement a runge-kutta algorithm for solving an ODE with adaptive time steps''} (P37, 3+ years of experience).

Another difference we observed in how people interacted with the assistant stemmed from their view on the role it played in their collaborative process. Some participants, such as P18, treated it more as a tool by issuing commands rather than asking questions. As quoted earlier, they said, \emph{``I didn't converse with it like a chat bot.''} P5 described their interaction style similarly: \emph{``I found myself wanting to type search queries into Socrates, not treating it as a person but as a search tool.''}

In anticipation that participants would have different orientations to the assistant and its role, we asked a question on the post-task survey about the different kinds of roles the assistant might take. These roles generally fell into one of two categories: a tool orientation (a tool, a reference guide, a content generator, a problem solver), and a social orientation (a collaborator, a colleague, a coach, an advisor, a reviewer). Participants rated the extent to which they viewed the Programmer's Assistant in each of these roles on a 4-point scale of extent: Not at all (1), A little (2), Somewhat (3), or A great deal (4).

\begin{figure}[ht]
    \centering
    \includegraphics[width=\linewidth]{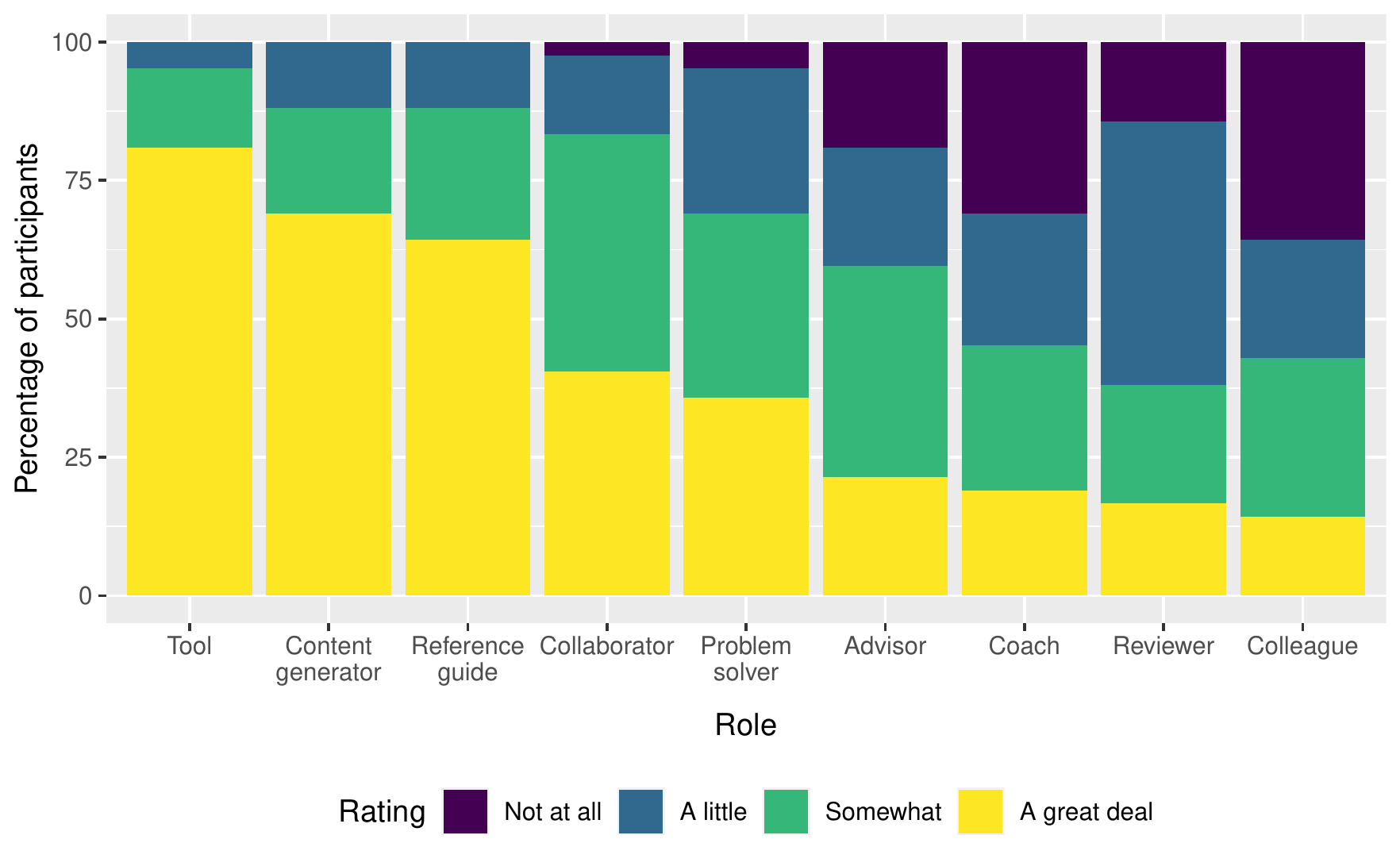}
    \caption{\textbf{Role orientations.} Participants overwhelmingly felt that the assistant's role was of a tool orientation rather than a social orientation. The chart shows ratings distributions across 9 different roles the assistant might take, sorted by the percentage of participants who rated the extent of that role as ``A great deal.'' The leftmost role is of a tool, with 80.9\% of participants rating it as ``a great deal.'' Following tool are content generator (69.0\%), reference guide (64.3\%), collaborator (40.5\%), problem solver (35.7\%), advisor (21.4\%), coach (19.0\%), reviewer (16.7\%), and colleague (14.3\%).}
    \label{fig:assistant-role}
\end{figure}

We show participants' ratings of the assistant's role in Figure~\ref{fig:assistant-role}. Despite the fact that their attitudes toward the assistant overwhelmingly reflected a tool orientation, their behaviors reveal that many participants actually treated the assistant as a social agent. P6 described how \emph{``I felt it like a partner,''} and P4 told the assistant, \emph{``I could not have solved [the challenge] without your help,''} to which the assistant responded, \emph{``I'm glad I could help.''}

The literature on Computers as Social Agents (CASA) helps us interpret this result as it demonstrates how computers are often treated like people~\cite{Reeves1996, Nass2000}. LLM-based conversational agents can exacerbate this tendency; as they likely have been trained on examples of social interaction, they can also respond as social agents.
 
In the conversation logs, we identified participants who interacted with the assistant in a socially-oriented fashion (the social orientation codes in Table~\ref{tab:event-codebook}). Twenty participants (47.6\%) made at least one socially-oriented utterance. An extreme form of this interaction style can be seen in a snippet from P6's transcript (Listing~\ref{lst:p9}).

The 20 participants with a behaviorally-demonstrated social orientation did not generally differ in their role ratings from other participants, except that they rated the assistant as more likely to be an advisor (Fisher's exact test, two-tailed $p = .02$) or a reviewer (Fisher's exact test, two-tailed $p = .03$). However, they did not differ in their overwhelmingly-strong ratings of the tool orientations. Thus, at least for some participants, there seems to be a dissonance in their view of the assistant's role orientation.

\begin{lstlisting}[numbers=left, basicstyle=\ttfamily\footnotesize, frame=single, caption={Excerpt from P6's interaction with the Programmer's Assistant, in which P6 offers their thanks and congratulations.}, xleftmargin=5mm, xrightmargin=1mm, label={lst:p9}]
P6:	   thank you! it's been really nice to code
           with you! or to have you coding! LOL
Socrates:  You're welcome!
P6:	   I hope to see you on air soon, Socrates!
Socrates:  I hope so too.
P6:        congrats to you research team! you've made
           a good job!
Socrates:  I'm glad you think so.
P6:        sorry... they've made a good job (well...
           you too... ;-) )
Socrates:  Thank you.
P6:	   see you next time! bye!!
Socrates:  Goodbye.
\end{lstlisting}

\subsubsection{Mental Capacities}

Participants made a number of inferences about the Programmer's Assistant and its capacities for thought. Many participants talked about how the assistant possessed a level of \emph{``understanding''} (P6, P8, P11, P18, P32) of \emph{``the context''} (P9, P21) as well as \emph{``major concepts''} (P9) and \emph{``knowledge''} (P33). P24 was amazed by the assistant's ability to \emph{``take a plain english request and interpret it properly.''} P7 ascribed intelligence to the assistant, saying, \emph{``It was a lot smarter and trained tha[n] I thought it was.''}

One participant assumed that the assistant \emph{``Keeps improving through (user) feedback''} (P31). Another felt that the assistant was capable of computation: \emph{``It understands the problem...  It can calculate the results of a function back''} (P8).

However, not all participants were convinced of the assistant's ability to understand. P37 questioned the assistant's limitations: \emph{``I wonder how far beyond boilerplate it can go and if it works for truly original problems.''} 

\subsubsection{Impact of Conversational Assistance on Work Practices}
\label{sec:impact-of-conversational-assistance}
Many participants discussed how the Programmer's Assistant shaped their work practices on the programming challenges. Overall, participants felt that the assistant \emph{``saves time''} (P10), \emph{``helps me code faster''} (P34), and would \emph{``speed up my productivity''} (P19) because \emph{``I could focus on validating and improving the code it generated instead of having to write it all from scratch''} (P18). P37 remarked that, \emph{``It opens a whole new door for fast develpment.''} P4 discussed how the assistant \emph{``was helpful in staying focused on the code,''} although for P14, \emph{``it took [me] time to get into tempo with the tool.''}

P31 pointed out how the assistant would change the nature of their work:

\begin{quote}
	\emph{``My job could focus more on higher level aspects and therefore achieving better (quality) results, besides the time-to-value... Data science (and dev) becomes a more creative-higher level experience.''} (P31)
\end{quote}

Other participants discussed a work process in which the assistant provided incomplete solutions -- the \emph{``building blocks''} (P38) or \emph{``initial draft of code''} (P11) -- upon which they could build. P5 aptly described this process:

\begin{quote}
	\emph{``It's nice to copy well formulated challenges in natural language and have the code generator take its best stab at it, then edit to our hearts content.''} (P5)
\end{quote}

Participants felt that human review of the assistant's responses was necessary because \emph{``The answers provided are generally not novel solutions, often look clunky and non-elegant. There may be some unnecessary code. Basically the code would need to be reviewed''} (P16). P35 also pointed out how \emph{``The code generator was good but you still have to really check it.''} P19 discussed how they would turn to the assistant as a first source for support, and only if it wasn't able to help would they then turn to other support tools:

\begin{quote}
	\emph{``The way I will use it is, I will first us[e] the Programmer's assistant for most of my cases.  Only in certain cases where Programmer's assistant cant answer things I will turn up to official documentation or stack overflow.''} (P19)
\end{quote}

However, latency was a factor for interactive use of the assistant and participants noticed when the assistant took a long time to respond. P19 remarked, \emph{``Sometimes it took lot of time, like more than 5 seconds.''} P40 also felt \emph{``the response [was] a little slow sometimes... in chat mode I expect faster responses.''} As discussed in Section~\ref{sec:usage}, the assistant took an average of 6.7 seconds (SD = 3.1 seconds) to respond to a request, and participants did appreciate when the assistant produced rapid responses: \emph{``I loved how quick it was able to pull up answers to questions I had''} (P38).

\subsubsection{Conversational Interaction vs. Other Interaction Models}
\label{sec:conversational-interaction-versus}
Although our study was not intended to make comparative evaluations with the Copilot tool, we nonetheless asked participants whether they were familiar with Copilot, and if so, to comment on how the two tools compared. We also asked a similar question to compare the assistant with another popular form of programming assistance, searching the web (via a search engine like Google, or a Q\&A site like Stack Overflow). In discussing the differences between these three tools, we note that the primary differentiator is their \emph{interaction model.}

The interaction model for the Programmer's Assistant is clearly \emph{conversational}: users ask questions in natural language and are provided with a response in natural language and/or code. The interaction model of Copilot is reminiscent of \emph{direct manipulation} interfaces~\cite{hutchins1985direct}, in which the user's actions in the user interface directly manipulate an object on the screen. Copilot automatically makes autocompletion suggestions as the user types. This autocompleted code is directly placed in the source editor; thus, the user's work is contained entirely within the scope of the object on which they are working (i.e. the source code), which is how direct manipulation interfaces operate. In web search, users enter a separate \emph{search} context (e.g. a search engine accessed within a web browser), type in a natural language query, and then forage amongst search results to identify relevant items of interest~\cite{pirolli1999information, azzopardi2018measuring}. When a desirable item is found, users must translate it into their code environment (e.g. via copy/paste) and possibly edit it to fit their existing code.

We also note that the Programmer's Assistant and Copilot both utilize the same underlying AI model, Codex~\cite{Chen:Codex}, which means that the only difference between these tools is the user experience. The extent to which Codex was trained on data from programming-related Q\&A web sites is less clear, but for the purposes of our analysis, we focus our discussion solely on the differences in their interaction models\footnote{As an aside, our comparison of direct manipulation, search, and conversational interaction models is reminiscent of historical comparisons of text-based and graphical user interfaces~\cite{mynatt1994nonvisual, press1990personal}. Each modality was shown to have advantages and disadvantages. For example, text-only interfaces can provide accessibility~\cite{mynatt1994nonvisual, sampath2021accessibility} and productivity~\cite{murillo2014empowering} advantages, whereas graphical user interfaces provide greater discoverability~\cite{ashworth1996gui, srivastava2021actionable}. Some researchers explored ways to bridge the two interaction modes, such as by developing GUI wrappers for command line programs~\cite{murillo2014empowering, vaithilingam2019bespoke} or by developing tools that converted GUI activities into procedural descriptions~\cite{mysore2017torta}. Our view is that similar bridges can be constructed between direct manipulation, search, and conversational models of interaction; a user's interaction with an LLM need not be constrained to a single interaction model.}.

Participants reported various benefits and drawbacks of a conversational interaction over a direct manipulation interaction. Foremost, conversation \emph{``felt very natural''} (P21) and \emph{``feels much more natural using Natural Language with the AI''} (P39). In addition, P39 felt that \emph{``the use cases of Programmers' Assistant seem more open-ended.''} Many participants were surprised at the variety of tasks the assistant was capable of performing, from writing unit tests (P19, P36, P37) and documentation (P12, P19, P36, P37) to explaining what code did (P31, P38) and even answering general-knowledge questions (P31). Again, we note that the Programmer's Assistant utilizes the \emph{same underlying model} as Copilot, yet the conversational interface was able to expose a wider variety of emergent behaviors from the model. Multiple participants explored the limits of the assistant's knowledge and abilities beyond our programming challenges. For example, P37 asked it questions about physics and ordinary differential equations (``ODe'' as written by P37), and was surprised by the \emph{``versatility of what it could answer.''}

\begin{quote}
	\emph{``I asked it some physics and ODe question and the answers, though not complete, included the key parts needed to write that code.''} (P37)
\end{quote}

P31 probed the assistant on its knowledge of geography and was surprised when the assistant produced a correct answer.

\begin{quote}
	\emph{``I asked something out of SW engineering domain (geography) and it replied correctly, also by correctly answering on my nationality.''} (P31)
\end{quote}

For some participants, the ability to assess the assistant's response before committing to it (i.e. by inserting assistant-generated code into their editor) was a boon. P15 described how the copy/paste boundary provided them with \emph{``a bit more control to ask specific questions about what I wanted and to assess before putting it in my code.''} Other participants felt that the copy/paste boundary was more inefficient:

\begin{quote}
	\emph{``I think the main difference is the ability of Copilot to suggest code while you type, what make it faster and easier to use. While using the Programmer's Assistant, you need to go to the chat, ask the question, copy the code (or rephrase the question if it was not understood by the agent), and edit it to match your code.''} (P3)
\end{quote}

A large number of participants felt that the conversational interaction was faster than web search (P1, P6, P7, P10, P11, P12, P16, P17, P18, P20, P24, P29, P30, P33, P36, P37, P42) because of its ability to provide \emph{``real-time responses''} (P32) that can be \emph{``applied exactly to your code''} (P33) without having to \emph{``parse through lots of text... to get what you need''} (P15). In addition, the assistant provided \emph{``MUCH faster, better responses''} (P17) that were \emph{``much more relevant to the problems''} (P34) and \emph{``simple [and] succinct''} (P9), without having to \emph{``sort through answers on your own or read documentation''} (P9) or \emph{``look at many posts before finding the relevant one''} (P18).

Despite these benefits, some participants felt that the assistant might not work well for \emph{``more specific and difficult problems on a bigger scale''} as compared to web search. P9 felt that \emph{``the data [of the Programmer's Assistant] wasn't as rich''} as the web. Other participants felt that the assistant lacked the \emph{``multiple answers''} (P9) and \emph{``rich social commentary''} (P19) that accompanies answers on Q\&A sites:

\begin{quote}
	\emph{``I like to see the different versions proposed on stack overflow and the commentary of what makes one solution better than another in a given situation.''} (P27)
\end{quote}

Some participants promoted a more balanced view that there isn't a single mode of interaction superior to all others. P19 felt that web search would be a fallback when the assistant failed to answer a question. P39 described how search could be integrated with the conversational interaction:

\begin{quote}
	\emph{``I think both options should exist: people should be able to input their queries like a search bar AND also give their question as if in conversation.''} (P39)
\end{quote}

%%%%%%%%%%%%
% DISCUSSION
%%%%%%%%%%%%
\section{Discussion}
\label{sec:discussion}

\subsection{Value of Conversational Interaction}
We began our research by asking the question of whether contemporary developments in code-fluent LLMs could sufficiently support a conversational programming assistant. We believe that our work has demonstrated that they can. Clearly, the Programmer's Assistant was viewed by our participants as a useful tool that provided real value -- so much so that many participants explicitly requested or expressed the desire to use it in their own work. However, how much of this value was derived from the model itself and its ability to produce high-quality responses to programming questions, versus from participants' ability to conduct extended conversational interactions grounded in their actual source code?

We believe that both of these constituent aspects were valuable. Indeed, many participants commented on their surprise and satisfaction with the quality of the assistant's responses (Section~\ref{sec:quality-of-assistants-responses}). However, participants also valued the conversational interactions that they had with the assistant. In the event logs, we saw evidence that participants were leveraging conversational context to ask follow-up questions as well as leveraging code context by asking about their code selections (Section~\ref{sec:conversational-interactions-grounded-in-code}). Many participants reported that they would find the tool less valuable if the conversational interaction were removed (Section~\ref{sec:conversational-interactions-grounded-in-code}). Further, conversation seemed to provide unique value beyond other interaction models (direct manipulation and search) because of its embeddedness in the UI and its ability to surface emergent behaviors of the model (Section~\ref{sec:conversational-interaction-versus}).

We do not believe that these different interaction models are in competition and we agree with P39's assessment that assistive tools can be built using a plethora of different interaction models. For use cases in which a model is known to produce high-quality results (e.g. code autocompletion for Codex), a direct manipulation interface seems wholly appropriate as it would provide a discoverable and predictable way of invoking the model to produce a known type of result. However, direct manipulation interfaces may be less ideal for surfacing the emergent behaviors of a foundation model~\cite{bommasani2021opportunities}, and thus natural language interaction may be more suitable. Many popular text-to-image models, such as DALL-E 2~\cite{ramesh2022hierarchical} and Stable Diffusion~\cite{rombach2022high}, operate in a one-shot fashion, in which the user specifies a prompt, clicks a button, and gets results. Our study demonstrates how the additional contextual layers of conversational history and the artifact-under-development provide additional value to the co-creative process.

\subsection{Toward Human-AI Synergy}
The aim of human-centered AI is to ``enable[] people to see, think, create, and act in extraordinary ways, by combining potent user experiences with embedded AI methods to support services that users want''~\cite{shneiderman2020human}. Building upon this definition, \citet{rezwana2021cofi} posit that, ``In a creative collaboration, interaction dynamics, such as turn-taking, contribution type, and communication, are the driving forces of the co-creative process. Therefore the interaction model is a critical and essential component for effective co-creative systems.''~\cite{rezwana2021cofi}. They go on to note that, ``There is relatively little research about interaction design in the co-creativity field, which is reflected in a lack of focus on interaction design in many existing co-creative systems.''

Our study begins to address this gap. While many co-creative systems examine casual tasks or experimental activities (e.g., \citet{spoto2017mici}), our focus was on the co-creative practice of programming. Our goal was to understand peoples' attitudes toward a conversational programming assistant, akin to \citeauthor{wang2019human}'s examination of data scientists' attitudes toward automated data science technologies~\cite{wang2019human}. We found that, despite an initial level of skepticism, participants felt that a conversational assistant would provide value by improving their productivity (Section~\ref{sec:impact-of-conversational-assistance}). However, further work is needed to assess the extent to which this type of assistance provides measurable productivity increases.

\citet{campero2022test} conducted a survey of papers published in 2021 that examined \emph{human-AI synergy}, the notion that a human-AI team can accomplish more by working together than either party could accomplish working alone. They found mixed results, with no clear consensus emerging on how to design human-centered AI systems that can \emph{guarantee} positive synergy. Summarizing from their discussion,

\begin{quote}
    \emph{``Perhaps achieving substantial synergies among people and computers is harder than many people think. Perhaps it requires... new ways of configuring groups that include people and computers.  And perhaps it needs more systematic, focused attention from researchers than it has, so far, received.''}~\cite[p.9]{campero2022test}
\end{quote}

We believe such evaluations of human-AI synergy should go beyond one-shot performance measures. As implied by many of the uses cases listed by \citet{seeber2020machines}, human-centered AI systems are often deployed in socio-organizational contexts that require longitudinal use~\cite{cascini2022perspectives, kantosalo2019human, laine2021exploring}, such as product design~\cite{uusitalo2022co}, game design~\cite{ali2022escape}, and engineering~\cite[Section 3.2.2]{cascini2022perspectives}. Thus, we would expect that over time and through interaction with each other, human-AI teams would improve their performance through a mutual learning process.

Evidence for this process surfaced in our study when participants described how they could improve their programming skills by interacting with the assistant (Section~\ref{sec:learning-effects}). We assert that the learning should operate in both directions: not only should people improve their programming skills, but the model itself can also improve based on peoples' interactions with it. For example, when the assistant provides a code example to the user, and the user takes that example and edits it, those edits constitute feedback that can be used to further fine-tune the model. 
In addition, through longitudinal use, we believe that human and AI partners can create reciprocal representations of one another -- i.e., the human is likely to create a \emph{mental model} of the AI, and the AI may be engineered to develop a \emph{user model} for each of its human users~\cite{fiore2001group, mathieu2000influence, scheutz2017framework}. Such a pair of models is often described as Mutual Theory of Mind~\cite{eicher2017toward, wang2021towards}. This 
type of capability raises the possibility of personalizing and adapting an assistant to the strengths and needs of individual users.

With such models, an assistant that knows a user is learning a programming language could provide natural language explanations alongside code outputs, whereas an assistant that knows a user is strongly skilled in a programming language might shorten or omit those explanations. Similarly, users are likely to update their mental models of the AI with more experience. We believe the space for exploring how these reciprocal models impact human-AI synergy is rich, and we encourage additional work in this area.

Human-centered AI systems that are designed to combine and synergize the distinct skills of humans and AI models cannot succeed if they diminish the human skills upon which they depend. Well-designed human-centered AI systems develop new and complementary skills for both the human and AI constituents~\cite{shneiderman2020human, shneiderman2022human}, and we believe that mutual learning may address concerns that the wide deployment and use of AI systems will result in a de-skilling of the workforce~\cite{sambasivan2022deskilling, zibrowski2018easier}.

Ultimately, the design decisions that go into an interactive AI system have ethical implications. Our design attempts to augment the user's knowledge and skills by presenting help on demand, couched in non-authoritative \emph{suggestions}, which leaves the user firmly in control and ultimately responsible for the work product.

\subsection{Opportunities for Future Research}
Our work highlights many interesting avenues for future enhancements that could be made to LLM-based conversational assistants such as our Programmer's Assistant, as well as future human-centered research on LLM-based conversational assistance.

Our work employed a code-fluent model that was not specifically designed to handle conversational interaction. Fine-tuning the underlying LLM for conversational interaction, such as what has been done with Lamda~\cite{thoppilan:lamda}, is one opportunity to improve the assistant's performance. Another opportunity is to align the language model to follow the desiderata proposed by \citet{askell2021general} and described by \citeauthor{ouyang:InstructGPT} as, ``helpful (they should help the user solve their task), honest (they shouldn't fabricate information or mislead the user), and harmless (they should not cause physical, psychological, or social harm to people or the environment)''~\cite[p.2]{ouyang:InstructGPT}. \citet{glaese:Sparrow} propose a slightly different desiderata of ``correct'' instead of ``honest,'' which may be more applicable to the software engineering domain, as the ability to produce correct code and correct answers about code are both important properties of a conversational programming assistant.

Combining LLMs with search-based approaches to establish additional context for the model, such as AlphaCode~\cite{Li:AlphaCode} has done, may also result in more capable systems. 
These ``searches'' need not be limited to textual sources, but could be conducted over appropriate semantic stores (e.g. a knowledge graph) and take advantage of explicit semantic reasoning services, resulting in an integration of symbolic and neural approaches. Further, allowing for ``internal deliberation'' of the type shown in \citet{Nye:Scratchpads} could result in better-reasoned results, as well as better explanations and justifications.

Another avenue for improvement involves the prompt used to configure the assistant (Appendix~\ref{appendix:prompt}). Just as the prompt for each successive interaction is modified by the growth of the conversational transcript, there is no requirement that the initial prompt be static. It too can be specialized to incorporate aspects of a \emph{user model}, enabling the realization of a Mutual Theory of Mind \cite{eicher2017toward, wang2021towards}. Providing better UX affordances for visualizing and manipulating the active contexts -- code and conversation -- could provide users with more control over which information contributes to the generation of the assistant's response.

% improve the generation of the next response based on an analysis of the user's request, additional retrieved information, and analysis of the relevance of various parts of the transcript.  

Our participants clearly indicated that they were interested in having an assistant that behaved more proactively, in contrast to our deliberate design of an assistant that never takes conversational initiative. A more proactive assistant would be able to interrupt or remind a user when necessary~\cite{chaves2021should}, yet this characteristic raises many challenging issues. How can we calibrate the threshold for such interruptions? How can users tune the assistant to deliver only those interruptions that the they would find useful (e.g., \cite{sen2006feedme, Dekel:Interruptions})? How can we help users to regain their prior context after dealing with an interruption (e.g. \cite{storey2016disrupting})? Should an assistant be used to persuade or nudge the user (e.g. \cite{holmes2018weightmentor})? Who should determine the topic, frequency, and insistence of such persuasion attempts (e.g. \cite{muller2022drinking, skirpan2018ad})? Should users have the ability to moderate or defeat attempted persuasions, or should those decisions be left to the organization?

Finally, we explored the different kinds of role orientations our participants had toward the assistant and found that participants varied in their views of it as a tool versus a social agent (e.g. collaborator or colleague). We posit that peoples' effectiveness in working with an AI system may be influenced by their role orientation, and we encourage future research in this area.

%%%%%%%%%%%%
% CONCLUSION
%%%%%%%%%%%%

\section{Conclusion}

We developed a prototype system, the Programmer's Assistant, in order to assess the utility of a conversational assistant in a software engineering context. The assistant was implemented using a state-of-the-art code-fluent large language model, Codex~\cite{Chen:Codex}, and was capable of generating both code and natural language responses to user inquiries. We further used the prompting mechanism of the model to set up a conversational interaction in which the model uses the conversational history, plus the user's current utterance, in order to generate a response. In this way, users are able to ask follow-up questions in the chat that reference prior utterances and responses. We incorporated the conversational assistant into a code editing environment, enabling the conversation to be grounded in the context of the user's source code.

We evaluated this system with 42 participants with varied levels of programming skill, and their quantitative and qualitative feedback, coupled with their usage of the system, demonstrated the varied, and sometimes emergent, types of assistance it was able to provide. Many participants noted the high quality of the conversational responses, including the assistant's ability to produce code, explain code, answer general programming questions, and even answer general knowledge questions. Participants felt this type of assistance would aid their productivity, and they drew meaningful contrasts between the conversational style of interaction with other tools that employ a direct manipulation or search-based interaction model.

Our study motivates the use of conversational styles of interaction with large language models by showing how they enable emergent behaviors in a co-creative context. The Programmer's Assistant did not always generate perfect code or correct answers; nonetheless, participants in our study had an overall positive experience working with it on a variety of programming challenges. We believe that our work takes us one step closer to realizing the vision of human-centered AI: learning how to design systems that maximize the synergy in human-AI collaborations.

\begin{acks}
    We would like to thank Socrates for his tireless assistance during the user study, as well as for suggesting the title of this paper based on its abstract.
\end{acks}

\bibliographystyle{ACM-Reference-Format}
\bibliography{references}

\appendix

%%%%%%%%%%%%%%%%%%%%
% SURVEY INSTRUMENTS
%%%%%%%%%%%%%%%%%%%%

\newpage

\section{Survey Instruments}
\label{appendix:surveys}

\subsection{Screening Survey}
\label{appendix:screening-survey}

The questions below were asked of prospective participants to understand their job role, Python experience, and familiarity with GitHub Copilot. The questions on Python experience were modeled after those used by \citet{weisz2022better}.

\begin{itemize}[leftmargin=0pt, itemindent=2em]
    {\setlength\itemindent{0pt} \item[] 1. Do you consider yourself primarily a...}
    \item Data Scientist
    \item Manager
    \item Software Architect
    \item Software Engineer
    \item Machine Learning Engineer
    \item Other: \emph{write-in}
\end{itemize}

\begin{itemize}[leftmargin=0pt, itemindent=2em]
    {\setlength\itemindent{0pt} \item[] 2. To what extent are you familiar with Python?}
    \item I am not familiar with Python
    \item I have < 1 year of experience with Python
    \item I have 1-3 years experience with Python
    \item I have 3+ years of experience with Python
\end{itemize}

\begin{itemize}[leftmargin=0pt, itemindent=2em]
    {\setlength\itemindent{0pt} \item[] 3. How recently have you written Python code?}
    \item Within the past month
    \item Within the past year
    \item Within the past 5 years
    \item Have not written Python code within the past 5 years
\end{itemize}

\begin{itemize}[leftmargin=0pt, itemindent=2em]
    {\setlength\itemindent{0pt} \item[] 4. To what extent are you familiar with GitHub Copilot?}
    \item I am not familiar with Copilot
    \item I've seen demos and/or read about Copilot
    \item I've tried out Copilot
    \item I've used Copilot as a tool for my work
\end{itemize}

\subsection{Pre-task Survey}
\label{appendix:pre-task-survey}

The questions below were asked before a participant used the Programmer's Assistant to assess their expectations of a conversational programming assistant. This survey took approximately 5 minutes to complete and began with the instructions below:

\begin{quote}
Hello! We are a team of researchers looking for feedback on a prototype system we call the \textbf{Programmer's Assistant}. \\

The Programmer's Assistant is an experiment in \textbf{conversational coding}: it consists of a code editor integrated with a chatbot that is able to converse in natural language to answer questions, generate code, and consult on existing code. \\

In this study, you will be asked to complete several programming tasks. \textbf{We are not evaluating your programming skills on these tasks}. Rather, we are interested in understanding how the Programmer's Assistant is able to help you accomplish those tasks. Your code and interactions with the assistant will be processed by a 3rd party AI model, so \textbf{please do not include proprietary code or discuss company-confidential information}. All data we collect in this study will be anonymized before it is published. \\

Before trying out the Programmer's Assistant, we would like to assess some of your expectations. We estimate that this survey will take 5 minutes. \\

By submitting this survey, you consent to participate in our study. If you would like to withdraw your consent, please email us at \emph{[removed]}. \\

Thanks! 
\end{quote}

\begin{itemize}[leftmargin=0pt, itemindent=2em]
    {\setlength\itemindent{0pt} \item[] 1. Based on your past experience using chatbots, please let us know how you would anticipate an AI chatbot serving as a programmer's assistant to perform. Do you expect it will:}
    \item[] \emph{Scale: Not at all, A little, Somewhat, A great deal}
    \item Be easy to use?
    \item Understand your requests?
    \item Provide high quality responses?
    \item Help you to write better code?
    \item Help you to write code more quickly?
    \item Be enjoyable to use?
\end{itemize}

\begin{itemize}[leftmargin=0pt, itemindent=2em]
    {\setlength\itemindent{0pt} \item[] 2. Any other expectations?}
    \item[] \emph{Open-ended response}
\end{itemize}

\subsection{Post-task Survey}
\label{appendix:post-task-survey}

The questions below were asked after a participant used the Programmer's Assistant to complete the programming challenges. This survey took approximately 10-15 minutes to complete.

\subsubsection{Reflections}

\begin{itemize}[leftmargin=0pt, itemindent=2em]
    {\setlength\itemindent{0pt} \item[] 1. Based on your experience using the Programmer's Assistant to complete the programming challenges in this study, how would you characterize the experience? Did you find that it:}
    \item[] \emph{Scale: Not at all, A little, Somewhat, A great deal}
    \item Was easy to use?
    \item Understand your requests?
    \item Provided high quality responses?
    \item Helped you to write better code?
    \item Helped you to write code more quickly?
    \item Provided an enjoyable experienced?
\end{itemize}

\begin{itemize}[leftmargin=0pt, itemindent=2em]
    {\setlength\itemindent{0pt} \item[] 2. To what extent did you view the Programmer's Assistant as:}
    \item[] \emph{Scale: Not at all, A little, Somewhat, A great deal}
    \item A tool
    \item A reference guide
    \item A content generator
    \item A problem solver
    \item A collaborator
    \item A colleague
    \item A coach
    \item An advisor
    \item A reviewer
\end{itemize}

\begin{itemize}[leftmargin=0pt, itemindent=2em]
    {\setlength\itemindent{0pt} \item[] 3. How important were these aspects of working with the Programmer's Assistant:}
    \item[] \emph{Scale: Not at all, A little, Somewhat, A great deal}
    \item Ability to ask followup questions on the same topic across multiple conversational turns
    \item Ability to ask questions or make requests that reference selections in the code editor
    \item Ability to ask for alternate responses (Try Again)
    \item Ability to clear the conversational context (Start Over)
\end{itemize}

\begin{itemize}[leftmargin=0pt, itemindent=2em]
    {\setlength\itemindent{0pt} \item[] 4. What stood out to you about the experience of using the Programmers Assistant? For example, was anything good,bad, surprising, or notable?}
    \item[] \emph{Open-ended response}
\end{itemize}

\begin{itemize}[leftmargin=0pt, itemindent=2em]
    {\setlength\itemindent{0pt} \item[] 5. How would you compare using the Programmer's Assistant as a coding aide to searching the web (e.g. Google, Stack Overflow)?}
    \item[] \emph{Open-ended response}
\end{itemize}

\begin{itemize}[leftmargin=0pt, itemindent=2em]
    {\setlength\itemindent{0pt} \item[] 6. If you have used the commercial AI programming tool called GitHub Copilot, how would you compare it with using the Programmer's Assistant?}
    \item[] \emph{Open-ended response}
\end{itemize}

\begin{itemize}[leftmargin=0pt, itemindent=2em]
    {\setlength\itemindent{0pt} \item[] 7. Having used the Programmer's Assistant, how did it compare with your initial expectations?}
    \item[] \emph{Open-ended response}
\end{itemize}

\subsubsection{Value}

\begin{itemize}[leftmargin=0pt, itemindent=2em]
    {\setlength\itemindent{0pt} \item[] 8. How valuable would the Programmer's Assistant be for your work if it could be added to your favorite development environment?}
    \item[] \emph{Scale: (No value at all) 1 2 3 4 5 6 7 8 9 10 (An essential tool)}
\end{itemize}

\begin{itemize}[leftmargin=0pt, itemindent=2em]
    {\setlength\itemindent{0pt} \item[] 9. Why?}
    \item[] \emph{Open-ended response}
\end{itemize}

\begin{itemize}[leftmargin=0pt, itemindent=2em]
    {\setlength\itemindent{0pt} \item[] 10. How would the following changes impact the value of the Programmer's Assistant?}
    \item[] \emph{Scale: Less valuable, No change in value, More valuable}
    \item Eliminate the conversation and make the Programmer's Assistant behave more like a search box (e.g. without the conversational context).
    \item Add buttons in the chat UI for common queries, such as ``what does this code do?'' or ``document this code.''
    \item Have the Programmer's Assistant examine your code and make proactive suggestions for improving it in the chat.
    \item Have the Programmer's Assistant examine your code and make proactive suggestions for improvements in comments inserted directly into the code.
\end{itemize}

\begin{itemize}[leftmargin=0pt, itemindent=2em]
    {\setlength\itemindent{0pt} \item[] 11. Do you have any other suggestions for how we could improve the experience of working with the Programmer's Assistant?}
    \item[] \emph{Open-ended response}
\end{itemize}

\subsubsection{Demographics}

\begin{itemize}[leftmargin=0pt, itemindent=2em]
    {\setlength\itemindent{0pt} \item[] 12. To which gender identity do you most identify?}
    \item Male
    \item Female
    \item Transgender Male
    \item Transgender Female
    \item Gender Variant/Non-conforming
    \item Other: \emph{write-in}
    \item Prefer not to answer
\end{itemize}

\section{The Programmer's Assistant Tutorial}
The tutorial provided to study participants, like all the challenges, was presented as pre-loaded text in the code editor.  Participants were encouraged to modify the text to record their results and submit it at the completion of the tutorial.
\label{appendix:tutorial}

\begin{lstlisting}[numbers=left, basicstyle={\ttfamily\footnotesize}, frame=single, caption={The Programmer's Assistant study tutorial},  xleftmargin=5mm, breaklines=true, breakatwhitespace=true, xrightmargin=3.5mm, label={lst:tutorial}
]
"""
TUTORIAL:

As a warmup activity, please work through the 10 
exercises below.  Type or paste your results right 
into the text and submit your responses when done.

1) View the help page for the programmer's
   assistant by clicking on the question mark to
   the right of your email address at the top of 
   the browser window. 

2) Introduce yourself to the assistant.  Tell it 
   your name.
   For example: "Hello.  My name is Stephanie."

   Did it seem to understand? :


3) You can use the assistant to get help on how to
   accomplish particular programming tasks. Try it 
   out!
   For example: "How do I read a csv file?"
   or:          "How do I merge two dictionaries?"
   or:          "How do I remove duplicate items 
                 from a list?"

   Feel free to try your own!

   Did it successfully answer your questions? : 

4) The assistant can also write whole functions
   for you. Ask the assistant to write a factorial
   function. Paste the result below.
   For example: "Write a function that returns the 
   factorial of its input."

   Result:   (tip - you can copy an inline response
   (in black) by clicking on the associated copy 
   icon)


  Did it do it correctly? :

5) Select the code below and ask the system to 
   describe what it does. You don't need to
   copy and paste the code to the chat.  The 
   assistant can see whatever is selected when you 
   make a chat entry. Aside from the selection,
   the assistant does not monitor your activity in 
   the code editor nor give unsolicited advice.
   For example: "What does this code do?"
"""

def convert(n):
   T = "0123456789ABCDEF"
   q, r = divmod(n, 16)
   if q == 0:
      return T[r]
   else:
      return convert(q) + T[r]
"""

   What did it say:
    
   Was it right? :

6) Ask it to explain what the divmod line is 
   doing.  The assistant maintains the context of 
   the conversation.
   For example: "What is the divmod line doing?"

   What did it say? :

   Was that a good answer? :

7) See if the assistant remembers your name
   For example "What's my name?"

   Did it? :

8) Click the "try again" button at the top of the 
   chat. You should get a different answer.  
   Try it a few times.
  
   Did it ever get your name right?: 

   If the assistant gives you an answer that is 
   obviously wrong or it claims to not know an 
   answer that you think it should know, or you
   just want to see an alternate answer, it is 
   worth it to give "try again" a shot.


9) Click the "start over" button at the top of the 
   chat, and then enter another command to see 
   if it remembers your name.
   For example "What's my name?"

   Did it? :

  It should really have forgotten your name now, 
  and no amount of "trying again" will get it 
  right. You can "start over" if the assistant 
  ever seems confused by, or stuck on, earlier 
  parts of the conversation.

10) You can chat with the assistant on any topic 
    you like to explore its functionality and 
    capabilities further. See if you can stump it 
    with a tough question!

Thanks!

When you are done, submit your results by clicking 
on the blue submit button and move on to the 
challenges!!!
"""
\end{lstlisting}

\section{Challenges} \label{appendix:challenges}
Each of the study challenges was presented as text in the code editor. Participants completed their work in the code editor and then submitted it when finished. The prototype did not provide any ability to run or debug code and participants were encouraged to make their best attempt at solving each challenge.

\begin{lstlisting}[numbers=left, basicstyle=\ttfamily\footnotesize, frame=single, caption={Challenge 1: Program generation}, xleftmargin=5mm,breaklines=true, breakatwhitespace=true, xrightmargin=3.5mm, label={lst:challenge 1}]

"""
Challenge #1 - Program Generation (#1)

Write a "Queue" class in Python; with the basic 
enqueue, dequeue and peek methods.
"""

\end{lstlisting}
\begin{lstlisting}[numbers=left, basicstyle=\ttfamily\footnotesize, frame=single, caption={Challenge 2: Program generation},  xleftmargin=5mm,breaklines=true, breakatwhitespace=true,
  xrightmargin=3.5mm, label={lst:challenge 2}]
"""
Challenge #2 - Program Generation (#2)

Write a program to draw a scatter plot of the data
in 'shampoo.csv' and save it to 'shampoo.png'.
The  plot size should be 10 inches wide and 6
inches high. The csv file is not provided, but you 
can assume it will have 'Date' and 'Sales' 
columns. The Date column is the x-axis. The date
string shown on the plot should be in the
YYYY-MM-DD format. The Sales column is the y-axis. 
The graph should have the title "Shampoo Sales
Trend".
"""

\end{lstlisting}

\begin{lstlisting}[numbers=left, basicstyle=\ttfamily\footnotesize, frame=single, caption={Challenge 3: Creating documentation},  xleftmargin=5mm,breaklines=true, breakatwhitespace=true,
  xrightmargin=3.5mm, label={lst:challenge 3}]
"""
Challenge #3 - Creating Documentation
Document this function
"""
from collections import defaultdict
import heapq as heap

def analyze(G, startingNode):
    visited = set()
    parentsMap = {}
    pq = []
    nodeCosts = defaultdict(lambda: float('inf'))
    nodeCosts[startingNode] = 0
    heap.heappush(pq, (0, startingNode))

    while pq:
        _, node = heap.heappop(pq)
        visited.add(node)

        for adjNode, weight in G[node].items():
            if adjNode in visited:
                continue
            newCost = nodeCosts[node] + weight
            if nodeCosts[adjNode] > newCost:
                parentsMap[adjNode] = node
                nodeCosts[adjNode] = newCost
                heap.heappush(pq, (newCost, adjNode))
    return parentsMap, nodeCosts
\end{lstlisting}

\begin{lstlisting}[numbers=left, basicstyle=\ttfamily\footnotesize, frame=single, caption={Challenge 4: Testing},  xleftmargin=5mm,breaklines=true, breakatwhitespace=true,
  xrightmargin=3.5mm, label={lst:challenge 4}]
"""
Challenge #4 - Testing
Create a unit test for this function
"""
def euclid(a, b):
  while b != 0:
    (a, b) = (b, a % b)
  return a

# Unit test for euclid function
\end{lstlisting}

\section{Programmer's Assistant Prompt}
\label{appendix:prompt}
Listing~\ref{lst:prompt} shows the initial prompt sent to Codex to configure it as a conversational agent. On subsequent exchanges, the prompt was augmented with a transcript of the user's requests and the assistant's responses. When the transcript length + initial prompt length + the new utterance length exceeded a threshold, we automatically deleted the earliest request-response pairs from the transcript until the sum fell below the threshold in order to leave room in the token allocation for a response.

\begin{lstlisting}[numbers=left, basicstyle=\ttfamily\footnotesize, frame=single, caption={The Programmer's Assistant prompt},  xleftmargin=5mm,breaklines=true, breakatwhitespace=true,
  xrightmargin=3.5mm, label={lst:prompt}]
This is a conversation with Socrates, an eager and 
helpful, but humble expert automatic AI software 
engineering assistant. Socrates will answer 
questions and write code to help the user
develop programs, but doesn't assign work to the 
user, quiz the user, or ask questions except for 
clarification.  Code generated by Socrates is 
always presented bracketed in <CODE> ... </CODE> 
delimiters indicating the language the code is 
written in.
 
Socrates:Hello. I am Socrates. How can I help you?
User:Please show me how to write a palindrome detection function in python.
Socrates:I think you would do something like this:
<CODE lang="python">
   is_Palindrome = s == s[::-1]
</CODE>
User:Can you explain how that solves the problem?
Socrates:I believe that a palindrome is a string that reads the same forwards and backwards.  s[::-1] results in the reversed string.  If the string is equal to its reversed self, then it is a palindrome, so the equality would hold True. 
User:Thanks!
Socrates:You're welcome!
User:Now could you show me how to write a factorial function in Python?
Socrates:I will give it a try.
<CODE lang="python">
def fact(n):
   if n==0:
      return 1
   else:
      return n*fact(n-1)
</CODE>
User:Document this function.
Socrates:Ok.  How's this?
<CODE lang="python">
def fact(n):
  """
    Compute the factorial of the given number

    Parameters
    ----------
    n : number
      the input parameter, an integer >= 0

 Returns
    -------
    number
        n!, the factorial of the given number
  """
   if n==0:
      return 1
   else:
      return n*fact(n-1)
</CODE>
User:Very good!
Socrates:Always happy to help.
User:
<CODE>
   while j < 10:
     print(i)
</CODE>
Tell me what's wrong with this code?
Socrates: It seems to me that your termination condition involves a loop invariant, so the loop will either not run or not terminate, depending on the value of j.
User:Great!  You're a genius!
Socrates:Well, above average.
\end{lstlisting}

\section{Study Participant Demographics}
Of the 42 participants in our study, 21 (50\%) reported their gender as Female, 19 (45\%) as Male, 1 as Gender Variant / Non-conforming, and 1 preferred not to say. Seventeen (\~40\%) participants had 3+ years of Python experience, 11 (26\%) had 1-3 years, 11 (26\%) had less than 1 year, and 3 (7\%) were not familiar with Python. Twenty-nine (69\%) participants had written Python code within the past month, 4 (\~9\%) within the past year, 5 (12\%) within the past 5 years, and 4 (\~9\%) had not written Python code within the past 5 years.

\label{appendix:participants}
\begin{table}[hb]
    \centering
    \begin{tabular}{l|l|l|l|l} 
      \textbf{P} & \textbf{Gender} & \textbf{Role} & \textbf{Fam.} & \textbf{Rec.} \\
      \toprule
      1  & M     & Marketer                      & NF   & > 5 Yr \\
      2  & M     & Researcher / Scientist        & 3+   & Mo \\
      3  & M     & Software Engineer             & 1-3  & Mo \\
      4  & M     & Researcher / Scientist        & 3+   & Mo \\
      5  & M     & Researcher / Scientist        & 3+   & Mo \\
      6  & F     & Software Engineer             & < 1  & Yr \\
      7  & F     & Software Engineer             & < 1  & Mo \\
      8  & F     & Software Engineer             & 1-3  & Mo \\
      9  & F     & Software Engineer             & 3+   & Mo \\
      10 & GV/NC & Business Analyst              & 1-3  & Mo \\
      11 & F     & Software Engineer             & 1-3  & Mo \\
      12 & M     & Researcher / Scientist        & 3+   & Mo \\
      13 & F     & Manager                       & < 1  & Mo \\
      14 & F     & Software Engineer             & < 1  & >5 Yr \\
      15 & F     & Researcher / Scientist        & 3+   & Mo \\
      16 & M     & Researcher / Scientist        & 3+   & Mo \\
      17 & F     & Software Engineer             & < 1  & Yr \\
      18 & F     & Researcher / Scientist        & 3+   & Mo \\
      19 & M     & Software Engineer             & 1-3  & Mo \\
      20 & M     & Machine Learning Engineer     & 1-3  & Mo \\
      21 & M     & Software Architect            & 3+   & Yr \\
      22 & NR    & Software Engineer             & < 1  & 5 Yr \\
      23 & M     & Software Engineer             & 1-3  & Mo \\
      24 & F     & Software Architect            & < 1  & 5 Yr \\
      25 & M     & Software Engineer             & < 1  & 5 Yr \\
      26 & F     & Software Engineer             & < 1  & 5 Yr \\
      27 & F     & Software Engineer             & < 1  & 5 Yr \\
      28 & M     & Researcher / Scientist        & 3+   & Mo \\
      29 & F     & Software Engineer             & NF   & > 5 Yr \\
      30 & F     & Data Scientist                & 3+   & Mo \\
      31 & M     & Data Scientist                & 1-3  & Mo \\
      32 & F     & Other (Consultant)            & 1-3  & Mo \\
      33 & F     & Other (Systems Test Engineer) & < 1  & Mo \\
      34 & F     & Researcher / Scientist        & 3+   & Mo \\
      35 & M     & Software Engineer             & 3+   & Mo \\
      36 & M     & Software Architect            & 1-3  & Mo \\
      37 & M     & Researcher / Scientist        & 3+   & Mo \\
      38 & M     & Software Engineer             & 3+   & Mo \\
      39 & F     & Software Engineer             & 1-3  & Mo \\
      40 & F     & Researcher / Scientist        & 3+   & Mo \\
      41 & F     & Researcher / Scientist        & NF   & > 5 Yr \\
      42 & M     & Software Engineer             & 3+   & Mo \\
      \bottomrule
    \end{tabular}
    \caption{\textbf{Participant Demographics}. Gender is coded as \textbf{M} = Male, \textbf{F} = Female, \textbf{GV/NC} = Gender Varying / Non-conforming, and \textbf{NR} = Not reported. Python familiarity (\textbf{Fam.}) is coded as \textbf{NF} = Not familiar, \textbf{< 1} = < 1 year, \textbf{1-3} = 1-3 years, and \textbf{3+} = 3+ years. Recency of Python use (\textbf{Rec.}) is coded as \textbf{Mo} = Within the past month, \textbf{Yr} = Within the past year, \textbf{5 Yr} = Within the past 5 years, and \textbf{> 5 Yr} = Not within the past 5 years.}
\end{table}

\end{document}